\documentclass[11pt]{article}

\usepackage[final]{acl}
\usepackage{times}
\usepackage{latexsym}
\usepackage{enumitem}
\usepackage[most]{tcolorbox}
\usepackage{booktabs}
\usepackage{array}
\usepackage[table]{xcolor}
\usepackage{enumitem}
\usepackage[T1]{fontenc}

\usepackage[utf8]{inputenc}

\usepackage{microtype}
\usepackage{booktabs}
\usepackage{multirow}
\usepackage{inconsolata}
\usepackage{amsmath}
\usepackage{graphicx}
\usepackage{subcaption}
\usepackage{pifont}
\usepackage{xcolor}
\definecolor{cFix}{HTML}{1B7F3A}     
\definecolor{cReg}{HTML}{D97706}     
\definecolor{cFail}{HTML}{C22B2B}    
\usepackage{booktabs}
\usepackage{array}
\usepackage{amssymb}
\newcommand{\ci}[1]{{\scriptsize$_{\pm\text{#1}}$}}
\definecolor{cPres}{HTML}{2563EB}    
\definecolor{cFunc}{HTML}{16A34A}    
\definecolor{cRob} {HTML}{DC2626}    
\newcommand{\chip}[2]{%
  \tikz[baseline=(c.base)]{\node(c)[
    inner xsep=4pt,inner ysep=1pt,rounded corners=2pt,
    fill=#1!12,draw=#1!55,line width=0.3pt,
    text=#1!85!black,font=\scriptsize\sffamily]{#2};}}
\newcommand{\tPres}{\chip{cPres}{Presence}}
\newcommand{\tFunc}{\chip{cFunc}{Functionality}}
\newcommand{\tRob} {\chip{cRob}{Robustness}}
\newcommand{\cn}[1]{\textbf{\textcolor{black!65}{(#1)}}}
\title{Asuka-Bench: Benchmarking Code Agents on Underspecified User Intent and Multi-Round Refinement}

\author{
 \textbf{Xin Wang\textsuperscript{1\textasteriskcentered}},
 \textbf{Liangtai Sun\textsuperscript{2,\textasteriskcentered}},
 \textbf{Yaoming Zhu\textsuperscript{2}},
 \textbf{Shuang Zhou\textsuperscript{2}},
 \textbf{Jiaxing Liu\textsuperscript{2}},
\\
 \textbf{Fengjiao Chen\textsuperscript{2}},
 \textbf{Lin Qiu\textsuperscript{2}},
 \textbf{Xuezhi Cao\textsuperscript{2}},
 \textbf{Xunliang Cai\textsuperscript{2}},
 \textbf{Licheng Zhang\textsuperscript{1}},
 \textbf{Zhendong Mao\textsuperscript{1, \textdagger}}
\\
\\
 \textsuperscript{1}University of Science and Technology of China
 \textsuperscript{2}Independent researchers
\\
 \textsuperscript{\textasteriskcentered}Equally contributed authors,
 \textsuperscript{\textdagger}Corresponding author
}

\begin{document}

\maketitle

\begin{abstract}
    Existing code-generation benchmarks score a single mapping from a complete prompt to a one-shot output. However, real web development is different. Users seldom write a full spec at the start; many requirements only become clear once they look at an intermediate result and react to it. We present \textbf{Asuka-Bench}, a benchmark that pairs underspecified user intent with multi-round refinement, grounded in browser-rendered behavior. Each task is resolved through a closed loop: a Code Agent generates a web project, a UI Agent executes test cases on the deployed site, and a User LLM turns evaluation outcomes into natural-language feedback for the next round. The benchmark comprises 50 web tasks with 784 evaluation criteria and 2{,}402 expected outcomes. We benchmark 8 LLMs across 2 agent frameworks. The results separate models clearly: weighted Task Pass Rate varies by 38 percentage points and models also differ substantially in their ability to repair from feedback. Asuka-Bench is also far from saturated: even the strongest model completes only 52\% of projects after three rounds.
\end{abstract}

\section{Introduction}

Large Language Models (LLMs) have shown remarkable progress in code generation, giving rise to increasingly capable code agents that can produce functional software from natural language instructions. To measure this progress, the community has developed a range of benchmarks spanning code completion~\cite{austin2021program,chen2021evaluating}, bug fix~\cite{jimenez2023swe,yang2024swe}, and end-to-end project generation~\cite{fu2025automatically,lu2025webgen}. Despite their diversity in task format, these benchmarks share a common paradigm: the input is a single, complete, and unambiguous specification, and the agent is expected to produce a correct output in one pass.

\begin{figure}
    \centering
    \includegraphics[width=0.95\linewidth]{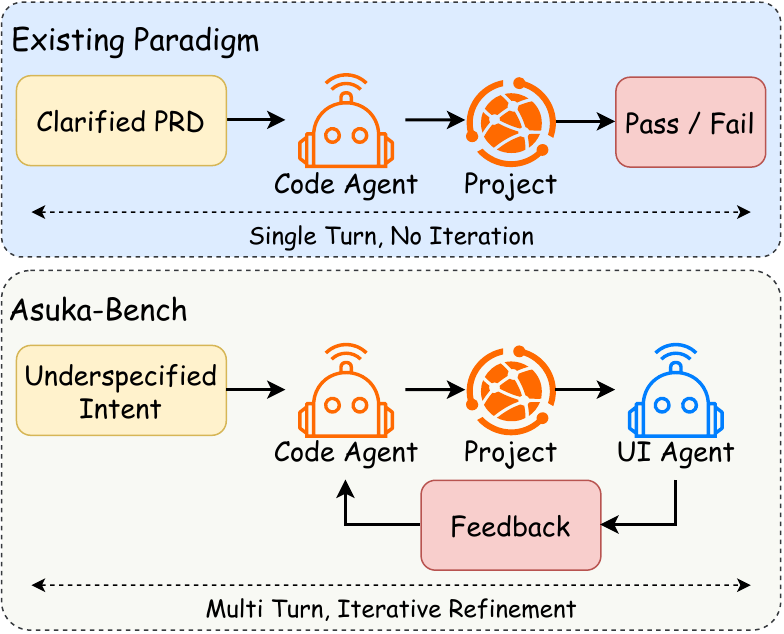}
    \caption{Comparison between existing paradigm and Asuka-Bench. Existing paradigm (top) uses clarified specs and single-turn execution; Asuka-Bench (bottom) supports iterative refinement from underspecified intents via agent feedback loops.}
    \label{fig:intro}
\end{figure}
This static instruction, one-shot output paradigm, however, fundamentally mismatches real-world development: users rarely articulate a complete specification upfront, and true requirements surface only as they inspect intermediate outputs, identify gaps, and provide corrective feedback. What ultimately distinguishes a capable code agent is therefore not whether it hits a full specification in one pass, but whether it converges under a feedback-driven loop. Web development is a uniquely natural testbed for studying this loop: web applications follow a what-you-see-is-what-you-get paradigm, allowing users to verify functional correctness, interaction logic, and error handling directly from the rendered interface without examining source code. This interaction-driven feedback loop is inherently absent in non-UI software such as command-line tools or libraries. Existing benchmarks, however, overlook this iterative process. By measuring only static, one-shot mappings from a complete prompt to a final output, they cannot evaluate whether an agent can incorporate execution feedback to progressively refine its work. This ability to iterate based on runtime feedback is essential for real-world deployment.

We close this gap with \textbf{Asuka-Bench}, a benchmark that pairs \emph{underspecified user intent} with \emph{multi-round iterative refinement}, and grounds both in the browser-observable behavior of the generated website. Each task begins with a deliberately underspecified request that mirrors how real users initiate development, and is resolved through a closed loop: (1)~a \textbf{Code Agent} generates an initial web project from the underspecified request; (2)~a \textbf{User Agent} deploys the project in a browser and drives an autonomous web-navigation module to exercise pre-defined test cases, judging outcomes from rendered behavior rather than source code; and (3)~the User Agent then synthesizes the per-criterion results into structured natural-language feedback. The Code Agent refines its implementation accordingly, and the loop repeats until all requirements are satisfied or a maximum number of rounds is reached.

Asuka-Bench comprises 50 web development tasks, each paired with a set of natural-language test cases that are executed through browser interaction, covering element existence, functional correctness, and robustness. We evaluate eight state-of-the-art LLMs across two agent frameworks. Results show that Asuka-Bench effectively differentiates model performance along two dimensions: weighted Task Pass Rate varies by more than 30 percentage points (pp), and models exhibit substantial differences in leveraging user feedback to fix errors over multiple rounds. The benchmark is also far from saturated: even the strongest model completes only 52\% of projects after three rounds.

Our contributions are summarized as follows:
\begin{itemize}
    \item \textbf{Benchmark.} We present Asuka-Bench, the first web generation benchmark that systematically evaluates code agents under underspecified user requests with multi-round interactive refinement, addressing a critical blind spot in existing static-instruction benchmarks.
    \item \textbf{An Evaluation Protocol Centered on User Feedback.} We ground evaluation in the user-feedback loop along two axes: \emph{interaction-as-feedback}, judging each criterion from rendered browser behavior rather than source code; and \emph{DAG-aware iterative evaluation}, which runs criteria in topological order on a DAG of dependencies and stops early so feedback stays focused on root causes, yielding 10.5\,pp higher task fix rates on average while cutting evaluation token cost by 23--26\%.
    \item \textbf{Comprehensive Empirical Analysis.} We evaluate 8 state-of-the-art LLMs across 2 agent frameworks, providing insights into how intent ambiguity and iterative feedback affect web generation performance.
\end{itemize}

\section{Related Work}

\subsection{Code Generation Benchmarks}

Early benchmarks for LLM-based code generation, such as HumanEval~\cite{chen2021evaluating} and MBPP~\cite{austin2021program}, evaluate models on self-contained function-level tasks with pre-defined unit tests. These benchmarks provide fully specified docstrings and assess whether a model can produce a correct implementation in a single pass. While they have been instrumental in tracking progress on code synthesis, they focus on isolated, short-horizon tasks that do not reflect the complexity of real-world software engineering.

To bridge this gap, repository-level benchmarks have been proposed. SWE-Bench~\cite{jimenez2023swe} and its multimodal extension~\cite{yang2024swe} require agents to resolve real GitHub issues by editing existing codebases, testing whether models can localize bugs, understand cross-file dependencies, and produce targeted patches. In the web domain, Web-Bench~\cite{xu2025web} constructs 50 projects with 20 sequentially dependent tasks each, simulating incremental feature development within an existing codebase and evaluating agents with pre-defined unit tests. WebMMMU~\cite{awal2025webmmu} further introduces multimodal code editing tasks that require models to modify HTML/CSS/JavaScript given screenshots and edit requests. While these benchmarks move closer to realistic development workflows, they still assume that the task specification is complete and unambiguous, where the agent receives a fully described issue or edit request and is expected to produce a correct patch without further clarification. A parallel line of work studies iterative code generation from self or environment grounded feedback~\cite{madaan2023selfrefine,shinn2023reflexion}; Asuka-Bench instead grounds refinement in simulated \emph{user} feedback derived from browser-rendered behavior.

\begin{figure*}[t]
    \centering
    \includegraphics[width=0.99\linewidth]{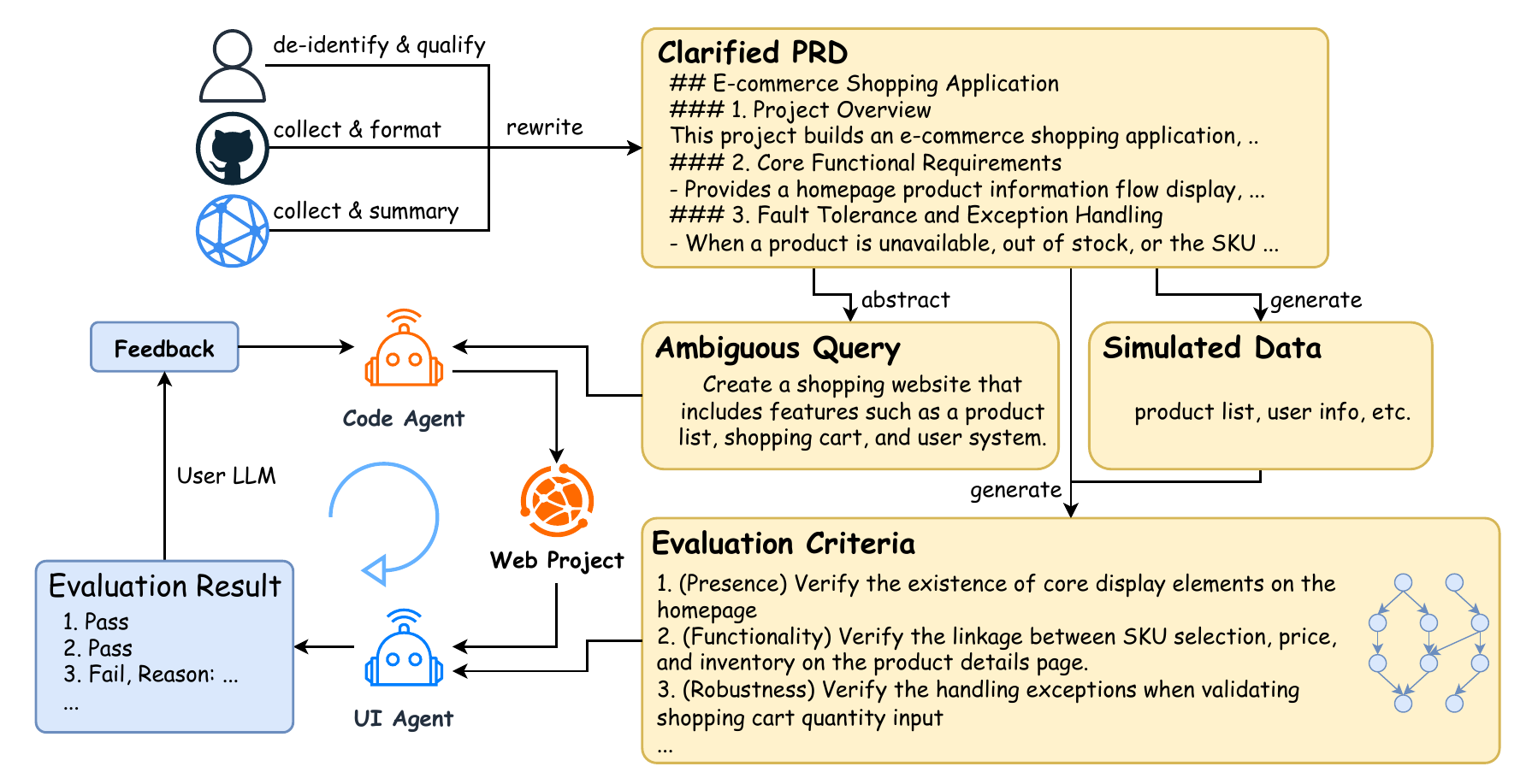}
    \caption{Overview of Asuka-Bench. \textbf{Top:} Dataset construction—raw queries from three sources are rewritten into a Clarified PRD, from which an underspecified query, simulated data, and DAG-organized evaluation criteria are derived. \textbf{Bottom-left:} Evaluation loop—a Code Agent generates a web project from the underspecified query; a UI Agent executes criteria as browser-based test cases; a User LLM synthesizes pass/fail results into feedback for the next refinement round.}
    \label{fig:framework}
\end{figure*}

\subsection{Web Generation Benchmarks}

A parallel line of work evaluates the ability of LLMs to generate complete web artifacts from scratch. One category focuses on \emph{screenshot-to-code} generation: given a webpage screenshot, the model must reproduce the corresponding HTML/CSS. Web2Code~\cite{yun2024web2code} and Design2Code~\cite{si2025design2code} evaluate this capability using visual similarity metrics, comparing the rendered output of the generated code against the original screenshot via structural, color, and layout dimensions. WebUIBench~\cite{lin2025webuibench} further introduces fine-grained element-level and layout-level evaluation by parsing DOM trees and computing matching scores. However, these benchmarks are inherently limited to visual fidelity and cannot assess functional correctness, user interaction logic, or robustness.

A more recent category evaluates \emph{text-to-web} generation: given a natural language description, the model must produce a deployable, interactive website. FrontendBench~\cite{zhu2025frontendbench} provides 148 prompt-test case pairs spanning five complexity levels and evaluates generated websites using Puppeteer-based unit tests. However, its prompts embed specific DOM identifiers (e.g., element \texttt{id} and \texttt{class} names) to facilitate automated testing, which deviates from realistic user specifications and introduces brittleness. WebGen-Bench~\cite{lu2025webgen} addresses this by using GPT-4o to generate natural-language test cases from task instructions and employing a web navigation agent (WebVoyager~\cite{he2024webvoyager}) to execute them on the generated websites, enabling more flexible and human-like evaluation. Despite this progress, all existing text-to-web benchmarks share two fundamental limitations: (1)~the input is a single, fully specified instruction that exhaustively describes the desired functionality, and (2)~evaluation is performed in a one-shot manner without iterative refinement. These assumptions are at odds with real-world web development, where users typically provide underspecified initial requests and progressively clarify their intent through dialogue.

\section{Asuka-Bench}

\subsection{Evaluation Framework}

Figure~\ref{fig:framework}~(bottom left) illustrates the overall evaluation pipeline of Asuka-Bench, comprising four stages: task specification, code generation, automated evaluation, and feedback-driven refinement.

\paragraph{Task Specification}
Each benchmark task originates from a deliberately \textbf{underspecified query} that mimics how real users initiate web development, e.g., ``Create a shopping website that includes features such as a product list, shopping cart, and user system.'' To establish ground-truth requirements without revealing them to the agent, we pair each underspecified query with a \textbf{Clarified PRD} (Product Requirements Document), a structured specification that decomposes the vague request into hierarchical functional modules with concrete requirements. The Clarified PRD covers not only core functional modules (e.g., product display with waterfall loading, category-based filtering, and price sorting) but also fault tolerance and exception handling rules (e.g., prohibiting cart additions and order placement when a product is unavailable or a SKU is out of stock). From each Clarified PRD, we further derive a set of \textbf{evaluation criteria}, i.e., natural-language test cases that describe expected behaviors at varying granularities, including element existence verification (``Verify the existence of core display elements on the homepage''), interaction logic validation (``Verify the linkage between SKU selection, price, and inventory on the product details page''), and exception handling verification (``Verify the handling of exceptions when validating shopping cart quantity input''). Crucially, only the underspecified query is provided to the Code Agent; the Clarified PRD and evaluation criteria serve exclusively as internal ground truth for automated assessment.

\paragraph{Code Agent}
Given the underspecified query, a Code Agent generates an initial web project. The Code Agent is instantiated with a backbone LLM and an agent framework that manages tool usage, file operations, and code execution. Across evaluation rounds, the Code Agent receives natural-language feedback from the User Agent and iteratively refines its implementation by modifying, adding, or restructuring project files.

\paragraph{Automated Evaluation}
Once the Code Agent produces a web project, the system deploys it in a browser environment. A \textbf{UI Agent}, which serves as an autonomous web-navigation module, then executes the pre-defined evaluation criteria as interactive test cases on the live website. For each criterion, the UI Agent performs the necessary browser interactions (navigating pages, clicking elements, filling forms, inspecting rendered content, etc.) and compares the observed behavior against the expected outcome specified in the criterion. Each test case yields a binary result: \emph{Pass} if the observed behavior matches the expectation, or \emph{Fail} accompanied by a natural-language explanation describing the discrepancy (e.g., ``the price did not update after selecting a different SKU'').

\paragraph{Feedback-Driven Iterative Refinement}
The per-criterion evaluation results are aggregated by a \textbf{User LLM}, which synthesizes them into structured natural-language feedback. This feedback is returned to the Code Agent, which uses it to diagnose failures and refine its implementation. The evaluation-feedback-refinement loop repeats until all evaluation criteria are satisfied or a pre-defined maximum number of interaction rounds is reached. This closed-loop mechanism enables Asuka-Bench to measure not only a code agent's initial generation quality but also its capacity to interpret underspecified feedback and iteratively converge on a correct implementation.

\subsection{Dataset Construction}

Figure~\ref{fig:framework}~(top right) illustrates the dataset construction pipeline. We describe the process in three stages: query sourcing, PRD construction, and evaluation criteria generation.

\begin{figure*}[t]
    \centering
    \begin{minipage}[c]{0.48\linewidth}
        \centering
        \includegraphics[width=\linewidth]{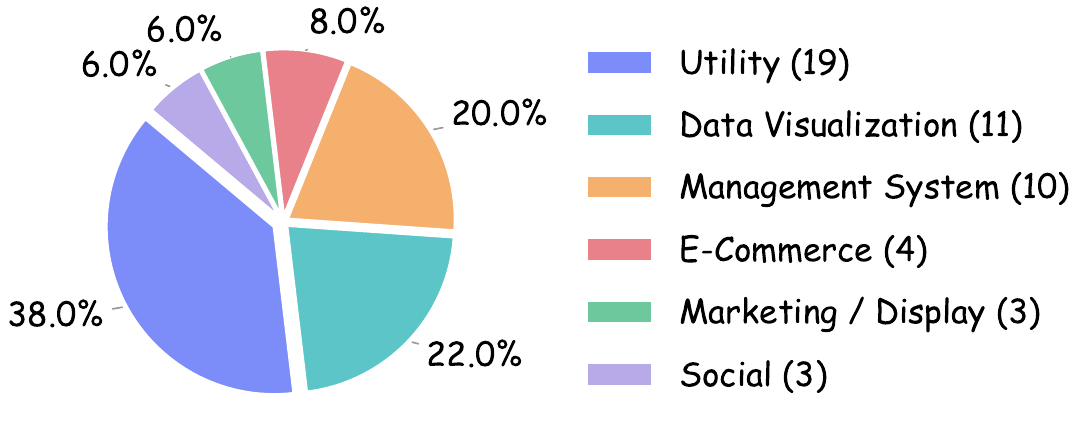}
    \end{minipage}
    \hfill
    \begin{minipage}[c]{0.48\linewidth}
        \centering
        \small
        \begin{tabular}{lr}
            \toprule
            \textbf{Statistic} & \textbf{Value} \\
            \midrule
            Number of tasks (PRDs) & 50 \\
            Application categories & 6 \\
            \midrule
            Total evaluation tasks & 784 \\
            \quad Presence & 129 (16.5\%) \\
            \quad Functionality & 492 (62.8\%) \\
            \quad Robustness & 163 (20.8\%) \\
            Total expected outcomes & 2{,}402 \\
            \midrule
            Avg.\ evaluation tasks / PRD & 15.7 \\
            Avg.\ expected outcomes / task & 3.1 \\
            Avg.\ difficulty weight & 2.96 \\
            \midrule
            Avg.\ DAG depth (levels) & 4.4 \\
            Max DAG depth & 9 \\
            \bottomrule
        \end{tabular}
    \end{minipage}
    \caption{Dataset overview. \textbf{Left:} Distribution of benchmark tasks across six application categories. \textbf{Right:} Summary statistics of the Asuka-Bench dataset.}
    \label{fig:dataset_stats}
\end{figure*}

\paragraph{Query Sourcing}
We collect seed queries from three complementary sources to ensure diversity in domain, complexity, and interaction pattern. (1)~\emph{Online user data.} We sample real user requests from a production web development service, apply de-identification to remove personally identifiable information, and filter for high-quality queries that involve non-trivial functional requirements. (2)~\emph{GitHub repositories.} We curate a set of high-quality open-source repositories and formalize their corresponding web applications into structured queries that describe the target functionality. (3)~\emph{Existing websites.} We select a diverse set of established, feature-rich websites (e.g., e-commerce platforms, knowledge forums) and summarize each into a query that captures its core functional characteristics. Together, these three sources yield a pool of raw queries spanning a broad range of web application scenarios.

\paragraph{Task Formalization}
We employ an LLM to rewrite each raw query into a Clarified PRD that decomposes the high-level intent into hierarchical functional modules with concrete specifications; see~\autoref{sec:synthesize} for details. From the Clarified PRD, we derive two additional artifacts. First, we abstract the PRD into a deliberately \textbf{underspecified query} by removing implementation details, specific interaction logic, and edge-case requirements, retaining only a high-level description of the desired application. This underspecified query serves as the sole input to the Code Agent during evaluation. Second, since each benchmark task requires a fully functional frontend application without backend dependencies, we generate simulated data (e.g., mock API responses, sample product catalogs, user profiles) based on the PRD to enable self-contained frontend execution.

\paragraph{Evaluation Criteria Generation}
The evaluation criteria constitute the core assessment instrument of Asuka-Bench. Given a Clarified PRD and its associated simulated data, we generate a structured set of evaluation criteria that the UI Agent will execute as test cases. The criteria are organized in a hierarchical structure: each PRD maps to multiple \textbf{evaluation tasks}, and each evaluation task comprises multiple \textbf{expected outcomes} that collectively determine whether the task is satisfied.

Each task is annotated with three attributes: \emph{task type}, \emph{difficulty weight}, and \emph{prerequisite dependencies}. Task types fall into three categories:
\begin{itemize}
    \item \textbf{Presence} tasks verify whether a required UI element or functional component exists in the generated application (e.g., ``Verify that the homepage contains a product search bar'').
    \item \textbf{Functionality} tasks validate whether an interactive feature behaves correctly according to its specification (e.g., ``Verify that selecting a different SKU updates the displayed price and inventory'').
    \item \textbf{Robustness} tasks assess whether the application handles edge cases and invalid inputs gracefully (e.g., ``Verify that adding a quantity exceeding available stock to the cart displays an appropriate error message'').
\end{itemize}
The difficulty weight assigns a scalar coefficient to each task, reflecting its implementation complexity, and is used to compute weighted scores during evaluation. Prerequisite dependencies encode logical ordering constraints: for instance, a functionality task that tests a feature's interactive behavior depends on the prior satisfaction of the presence task confirming that the feature's constituent elements exist. These dependencies induce a \textbf{directed acyclic graph (DAG)} over the evaluation tasks associated with each PRD.

\paragraph{DAG-Based Evaluation Protocol}
During evaluation, the UI Agent executes tasks following the topological order of the DAG. A task is executed if and only if all of its prerequisite dependencies have passed; if any prerequisite has failed or was not executed, the task is skipped without evaluation. After all reachable tasks have been processed, only the tasks that were actually executed and returned a failure are compiled into structured feedback and returned to the Code Agent for the next refinement iteration. Skipped tasks produce no feedback, as their failures are indirect consequences of upstream issues. This dependency-driven protocol avoids wasteful evaluation of tasks whose foundations are unmet, and focuses the Code Agent's attention on the root causes of failure rather than their downstream symptoms. We validate this design choice and analyze the evaluation gap introduced by dependency blocking in Appendix~\ref{sec:appendix_rq3}.

\begin{table*}[htbp]
  \centering
  \caption{Main results on Asuka-Bench. PCR: Project Completion Rate. R$_1$--R$_3$: cumulative weighted Task Pass Rate per round. Criteria: cumulative weighted Criteria Pass Rate at Round~3. 95\% CIs (CLT, $n{=}50$) shown as subscripts; best values per framework block in \textbf{bold}.}
  \label{tab:main_results}
  \setlength{\tabcolsep}{6pt}
  \begin{tabular}{ll *{5}{>{\centering\arraybackslash}p{11mm}}}
  \toprule
  \multirow{2}{*}{\textbf{Framework}} & \multirow{2}{*}{\textbf{Model}}
    & \multirow{2}{*}{\textbf{PCR}}
    & \multicolumn{3}{c}{\textbf{Cumulative Task Pass Rate}}
    & \multirow{2}{*}{\textbf{Criteria}} \\
  \cmidrule(lr){4-6}
  & & & R$_1$ & R$_2$ & R$_3$ & \\
  \midrule
  \multirow{8}{*}{OpenHands} & GPT-5.4 & \textbf{52} & \textbf{56.6\ci{7.3}} & \textbf{82.1\ci{5.6}} & \textbf{90.1\ci{4.7}} & \textbf{95.1\ci{2.6}} \\
   & Kimi-K2.6 & 50 & 44.3\ci{8.2} & 74.7\ci{7.8} & 86.6\ci{6.0} & 92.1\ci{4.8} \\
   & Gemini-3.1-Pro & 36 & 37.2\ci{7.9} & 69.2\ci{7.4} & 80.7\ci{6.8} & 89.4\ci{4.7} \\
   & GLM-5 & 32 & 40.5\ci{8.2} & 64.4\ci{8.8} & 75.3\ci{8.1} & 83.3\ci{7.2} \\
   & Qwen3.5-Plus & 24 & 40.2\ci{8.6} & 61.1\ci{9.0} & 72.9\ci{7.9} & 82.5\ci{6.4} \\
   & MiniMax-M2.7 & 24 & 36.7\ci{8.2} & 59.1\ci{9.2} & 70.7\ci{9.0} & 79.8\ci{8.1} \\
   & Seed-2.0-Pro & 14 & 26.9\ci{7.3} & 44.7\ci{9.3} & 53.7\ci{10.0} & 63.9\ci{9.7} \\
  \midrule
  \multirow{7}{*}{Claude Code} & Claude-4.6-Sonnet & \textbf{46} & \textbf{55.7\ci{7.5}} & \textbf{82.4\ci{6.2}} & \textbf{89.4\ci{5.7}} & \textbf{93.6\ci{5.4}} \\
   & GPT-5.4 & 44 & 49.7\ci{7.4} & 80.3\ci{6.4} & 88.1\ci{5.9} & 92.8\ci{4.8} \\
   & LongCat-Preview & 32 & 44.1\ci{7.8} & 66.8\ci{9.0}\ci{7.2} & 80.2\ci{6.8} & 86.3\ci{6.3} \\
   & GLM-5 & 30 & 40.8\ci{9.2} & 66.5\ci{9.2} & 76.6\ci{8.9} & 82.9\ci{8.5} \\
   & LongCat-2.0-Preview & 26 & 33.3\ci{9.1} & 61.3\ci{9.7} & 73.1\ci{9.1} & 80.0\ci{8.6} \\
   & Qwen3.5-Plus & 24 & 41.3\ci{8.3} & 60.4\ci{9.0} & 70.1\ci{8.3} & 80.2\ci{6.7} \\
   & MiniMax-M2.7 & 22 & 39.2\ci{8.3} & 65.7\ci{7.8} & 76.4\ci{6.6} & 86.3\ci{5.0} \\
   & Seed-2.0-Pro & 8 & 24.7\ci{8.0} & 39.8\ci{9.7} & 51.8\ci{10.1} & 60.9\ci{10.4} \\
  \bottomrule
  \end{tabular}
\end{table*}

\subsection{Dataset Statistics}

Asuka-Bench comprises 50 web development tasks spanning six application categories adopted from the ArtifactsBench taxonomy~\cite{zhang2025artifactsbench}. As shown in Figure~\ref{fig:dataset_stats}~(left), the distribution is dominated by Utility (38.0\%), Data Visualization (22.0\%), and Management System (20.0\%), followed by E-Commerce (8.0\%), Marketing/Display (6.0\%), and Social (6.0\%). This distribution is aligned with the real-world category distribution observed in our production user traffic, ensuring that the benchmark reflects practical development demand.

Figure~\ref{fig:dataset_stats}~(right) summarizes the key statistics. Across the 50 tasks, we derive a total of 784 evaluation tasks containing 2{,}402 expected outcomes, with an average of 15.7 evaluation tasks and 3.1 expected outcomes per task. In terms of task type, Functionality tasks account for the majority (62.8\%), followed by Robustness (20.8\%) and Presence (16.5\%), reflecting the emphasis on verifying interactive behaviors and edge-case handling beyond mere element existence. Difficulty weights range from 1 to 5, with an average weight of 2.96. The dependency-induced DAGs have an average depth of 4.4 levels (maximum 9), indicating that the evaluation criteria encode substantial inter-task dependencies that require systematic, layered verification. A complete walkthrough of one representative task is provided in ~\autoref{sec:fulldata}.

\section{Experiments and Results}

\subsection{Settings}
\label{subsection:settings}

\paragraph{Models and frameworks.}
We evaluate eight state-of-the-art LLMs as Code Agent backbones: GPT-5.4 (Medium)~\cite{openai2025gpt5}, Gemini-3.1-Pro~\cite{google2025gemini}, Claude-4.6-Sonnet~\cite{anthropic2025claude4}, GLM-5~\cite{glm2024chatglm}, Kimi-K2.6~\cite{kimi2025k2}, Seed-2.0-Pro~\cite{bytedance2025seed}, MiniMax-M2.7~\cite{minimax2025m27}, and Qwen3.5-Plus~\cite{qwen2025qwen3}. These LLMs are paired with two representative agent frameworks: OpenHands~\cite{wang2024opendevin}, an open-source sandbox-based agent with explicit tool calls, and Claude Code, a CLI-based agent with conversational tool orchestration. We report all accessible model--framework combinations, and use GPT-5.4 as the backbone for both the Evaluation Agent and the User Agent (feedback synthesis) across all configurations. Gemini-3.1-Pro and Kimi-K2.6 on Claude Code, and Claude-4.6-Sonnet on OpenHands are skipped, see~\autoref{sec:appendix_impl} for reasons.

\paragraph{Protocol and metrics.}
The maximum number of refinement rounds is 3; the UI Agent uses vision-enabled browser interaction with up to 100 steps per task. Tasks are evaluated under our DAG-aware protocol with soft-satisfaction threshold $T{=}0.5$; full hyperparameters are listed in~\autoref{sec:appendix_impl}. We report three metrics: \textbf{Project Completion Rate} (PCR, fraction of projects where all tasks pass), \textbf{cumulative weighted Task Pass Rate} ($\sum_{i \in \text{passed}} w_i / \sum_i w_i$ with difficulty weight $w_i$), and \textbf{cumulative weighted Criteria Pass Rate} ($\sum_i w_i (c_i^{\text{pass}}/c_i^{\text{total}}) / \sum_i w_i$), which softens the binary task indicator to the sub-criteria pass ratio.

\paragraph{Confidence intervals.}
For each (model, framework, round) cell, we report 95\% CIs via Central Limit Theorem normal approximation over the 50 per-project pass ratios: $\bar{x} \pm 1.96 \cdot s/\sqrt{50}$. Details are deferred to~\autoref{sec:appendix_impl}.

\subsection{Results}

Table~\ref{tab:main_results} reveals a clear separation across the 13 (model, framework) configurations: cumulative weighted Task Pass Rate after three rounds spans 51.8\%--90.1\%, a 38-point range with non-overlapping 95\% CIs at the top and bottom of the ranking.
The headline gap, however, is between tasks and projects: even the strongest configurations reach $\sim$90\% Task Pass Rate yet only 46--52\% Project Completion Rate, since a deployable web application requires \emph{all} of its specified tasks to pass. The residual failures concentrate on functional logic and robustness rather than on element presence, as detailed in the per-type breakdown of \autoref{sec:appendix_bytype}.
Iterative feedback contributes most of its gain in Round~2 ($\sim$25\,pp absolute), with Round~3 adding a smaller 7--13\,pp; no model saturates within three rounds.
A detailed framework comparison is provided in~\autoref{sec:appendix_framework}, and two qualitative case studies illustrate how round-wise feedback localizes and resolves cross-page defects in~\autoref{sec:appendix_case_study}.
 
\begin{figure}[htbp]
    \centering
    \includegraphics[width=\linewidth]{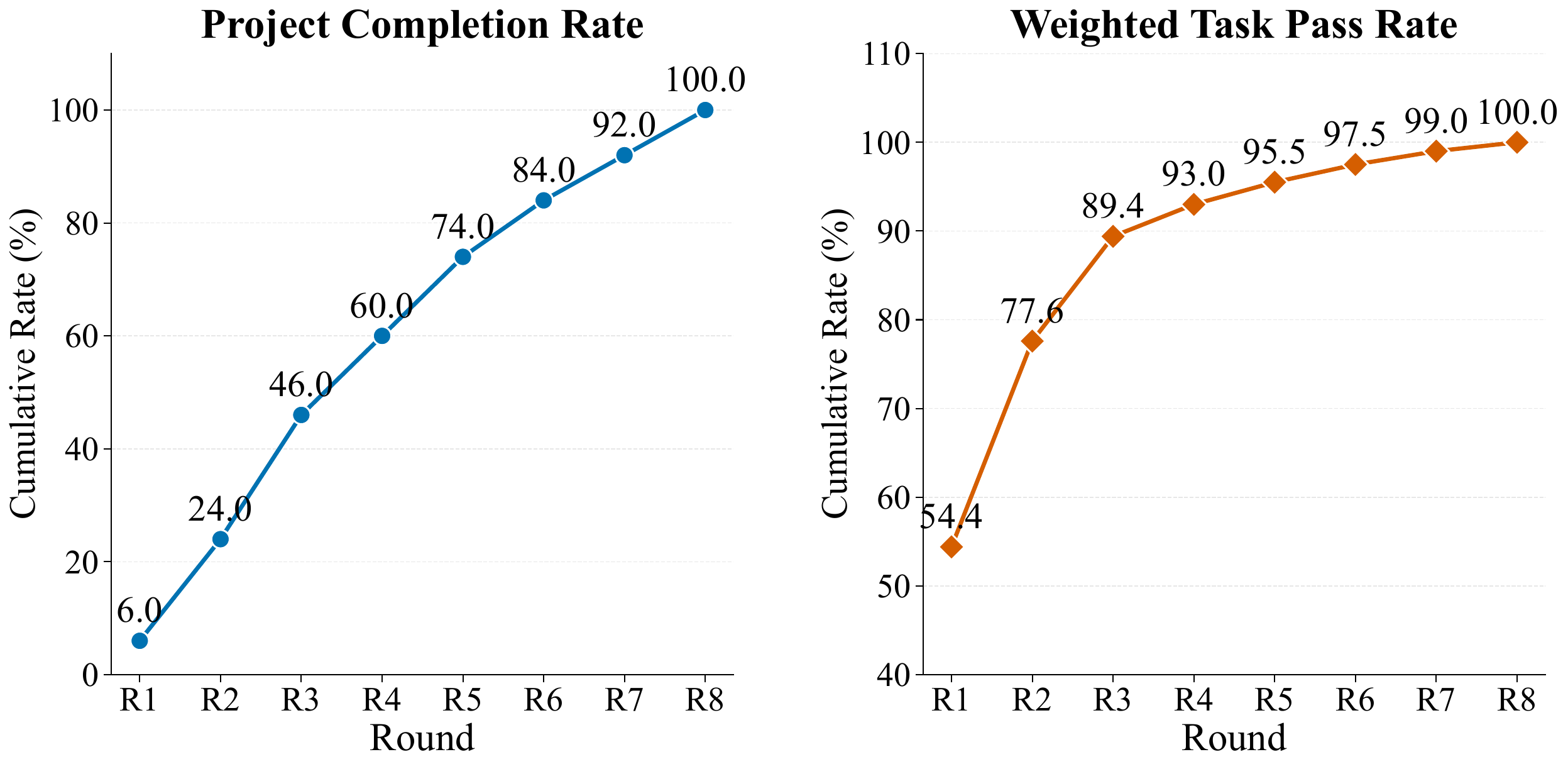}
    \caption{Cumulative Project Completion Rate and weighted Task Pass Rate for Claude-4.6-Sonnet across 8 feedback rounds, both reaching 100\%.}
    \label{fig:8round}
\end{figure}

\begin{figure}[htbp]
    \centering    \includegraphics[width=\linewidth]{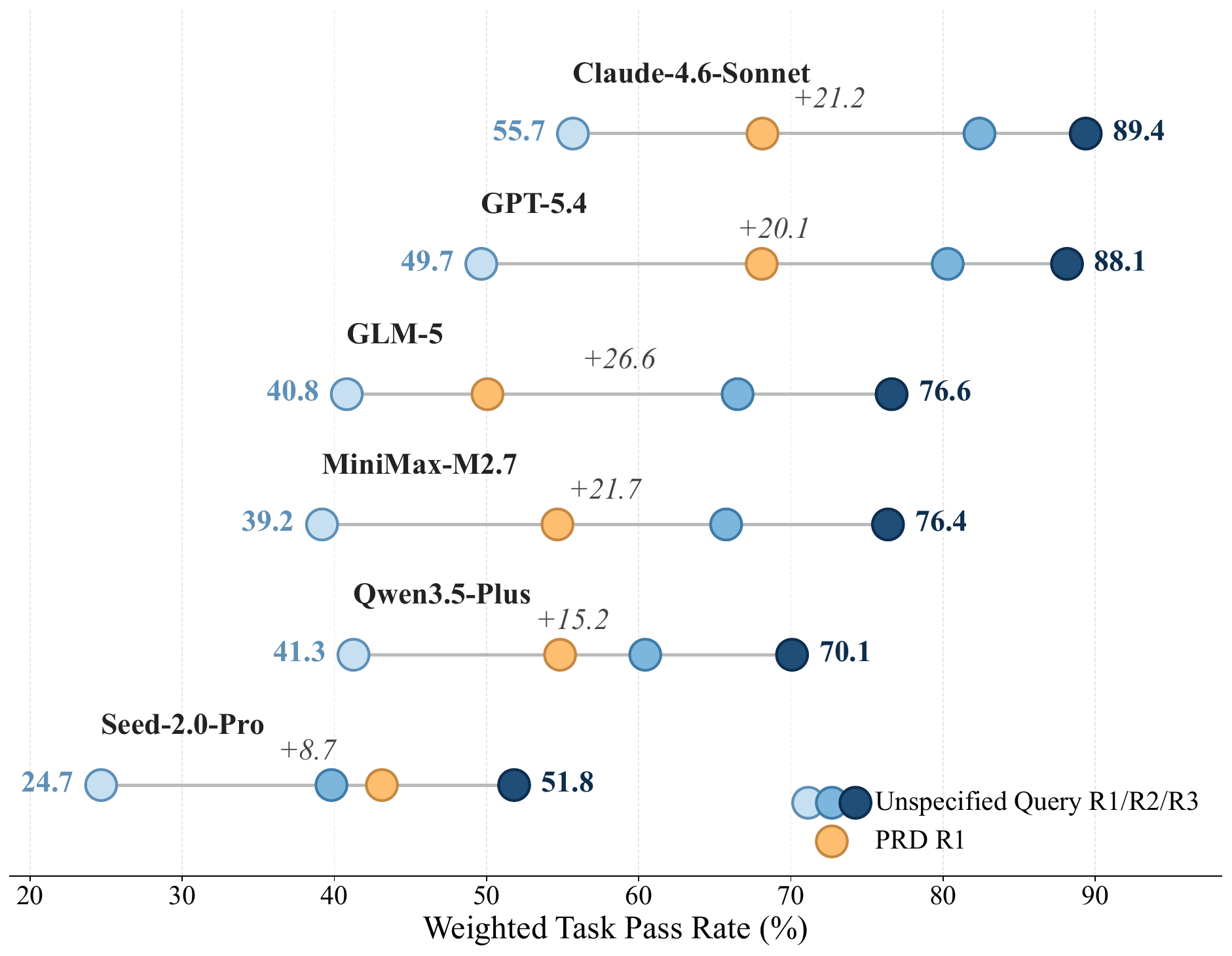}
    \caption{Weighted Task Pass Rate per Claude Code model under four settings: a single round on the Clarified PRD, and three Rounds on the underspecified query}
    \label{fig:rq1_bar}
\end{figure}

\paragraph{Saturation Analysis.} To verify that the evaluation tasks are inherently solvable rather than ill-defined, we extend the SOTA model (Claude-4.6-Sonnet) to 8 interaction rounds. As shown in Figure~\ref{fig:8round}, both Project Completion and weighted Task Pass Rate converge to 100\% by Round~8, confirming that all benchmark tasks are achievable given sufficient iterations. The monotonically increasing curve also demonstrates that our feedback mechanism consistently provides actionable guidance without introducing noise or regression.

\section{Analysis}
\begin{figure*}[htbp]
    \centering
    \includegraphics[width=0.9\linewidth]{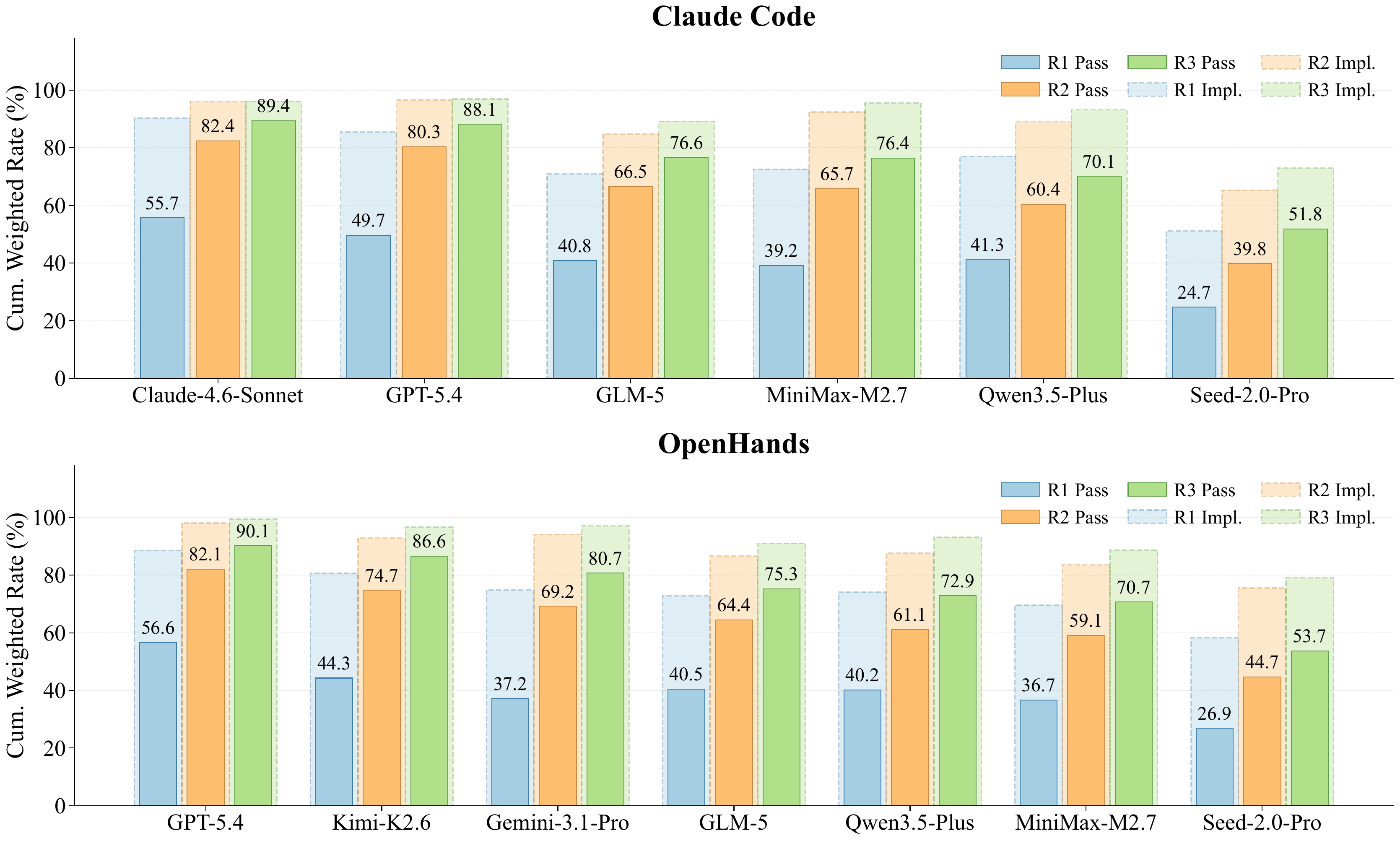}
    \caption{Cumulative weighted \textbf{Implemented} (translucent outer bar; tasks reachable under the DAG evaluation protocol) and \textbf{Pass} (solid inner bar; tasks fully passed) rates per round, under the underspecified-query setting. Top: Claude Code; bottom: OpenHands. The Implemented$-$Pass gap measures evaluable tasks the model failed to complete, and how it shrinks across rounds reflects each model's repair-from-feedback ability.}
    \label{fig:rq2_bar}
\end{figure*}

\paragraph{RQ1: Multi-round refinement on an underspecified query vs.\ a single-round Clarified PRD.}
We give each model two budgets that supply the same eventual information: a single round on the Clarified PRD (denoted PRD-1R), or three rounds on the underspecified query with browser-grounded feedback (Figure~\ref{fig:rq1_bar}, Claude Code, six models). The first ambiguous round loses 9--18\,pp relative to the PRD baseline, but every model overtakes its PRD score by Round~3 (net gains 8.7--26.6\,pp), confirming that multi-round refinement on an ambiguous start can match---and exceed---a fully specified one-shot prompt. The two protocols are also complementary. Under PRD-1R, Qwen3.5-Plus (54.8\%) edges out GLM-5 (50.1\%), and Claude-4.6-Sonnet ties GPT-5.4 (both 68.1\%); after three ambiguous rounds the order shifts (GLM-5: 76.6\% vs.\ Qwen3.5-Plus: 70.1\%; Sonnet: 89.4\% vs.\ GPT-5.4: 88.1\%). Asuka-Bench therefore complements existing one-shot benchmarks by surfacing repair-from-feedback ability, which we examine in RQ2.

\paragraph{RQ2: Is repair-from-feedback ability a separable dimension of model capability?}
Figure~\ref{fig:rq2_bar} pairs each round's Implemented rate (tasks the model attempted) with its Pass rate (tasks that fully passed); the gap between them is the share of attempted-but-broken tasks. Stronger models close it fast (GPT-5.4 on OpenHands: 99.5\% vs.\ 90.1\% at Round~3), while weaker models retain a persistent residual (Seed-2.0-Pro on OpenHands: 79.1\% vs.\ 53.7\%).

Crucially, first-round generation and later-round repair are \emph{not} the same axis: Kimi-K2.6 on OpenHands climbs 44.3\%\,$\to$\,74.7\%\,$\to$\,86.6\%, ending second-best in its framework after a mediocre first round. A single-round evaluation would have recorded only its 44.3\%, conflating ``cannot generate from an underspecified prompt'' with ``cannot benefit from feedback.'' Per-round fix rates are tabulated in \autoref{sec:appendix_fix_rate}.

\paragraph{RQ3: Does the closed-loop evaluation framework align with human judgment?}
To validate the reliability of our automated evaluation, we conduct a human-agreement study. We randomly sample 50 projects across five representative models (Claude-4.6-Sonnet, GLM-5, GPT-5.4, MiniMax-M2.7, Seed-2.0-Pro), covering 3{,}576 individual criterion-level annotations across the sampled projects. Expert annotators independently judge each criterion as Pass or Fail by interacting with the deployed web application; full annotation guidelines and the annotation console are provided in~\autoref{sec:appendix_annotation}. Taking human judgments as the gold standard, our BrowserUse-based evaluation agent achieves 93.35\% accuracy and 96.52\% F1 (precision 97.37\%, recall 95.68\%), demonstrating strong alignment with human assessment and confirming that the closed-loop evaluation framework provides a reliable, cost-effective proxy for human judgment in large-scale iterative benchmarking.

\section{Conclusion}
We presented Asuka-Bench, the first web generation benchmark that evaluates code agents under underspecified user intent with multi-round iterative refinement grounded in browser-observable behavior. Our evaluation of 8 state-of-the-art LLMs across 2 agent frameworks yields three main findings. First, intent ambiguity introduces a substantial performance gap, which iterative feedback effectively bridges within three rounds. Second, repair-from-feedback ability is largely decoupled from first-round generation quality, revealing a capability axis that existing one-shot benchmarks cannot surface. Third, our DAG-aware evaluation protocol provides reliable automated judgment. 
The benchmark is far from saturated, and we hope it encourages the community to move beyond static, one-shot evaluation toward benchmarks reflecting the iterative nature of real-world development.

\section*{Limitations}
Asuka-Bench has three main limitations that we wish to flag. First, in terms of \emph{domain coverage}, the 50 tasks span six categories sampled to mirror our production traffic and target self-contained frontends with simulated data; 3D, real-time collaborative, and live-backend scenarios remain out of scope and would require non-trivial extensions to both the data pipeline and the deployment harness. Second, regarding \emph{evaluator dependence}, both the UI Agent and the User LLM are instantiated with GPT-5.4: although our human-agreement study reports 93.35\% accuracy with single-annotator labels, swapping the evaluator backbone may shift absolute scores, and a multi-annotator follow-up would tighten the upper bound on automated-evaluator quality. Third, like any \emph{static benchmark surface}, Asuka-Bench's underspecified queries and criterion DAGs can in principle be memorized by future training runs; long-term contamination resistance would require periodic refresh of the task pool, which we leave to future work.

\bibliography{custom}

\appendix
\section{DAG-Based Evaluation Analysis}
\label{sec:appendix_rq3}

\paragraph{Why a soft satisfaction threshold.}
The DAG protocol skips a downstream task whenever its prerequisites are
deemed unmet. Exactly how strict that test should be is a design choice:
under \emph{strict} enforcement ($T{=}1.0$, every parent sub-criterion
must pass) a single peripheral defect in a parent blocks every
descendant; under \emph{soft} enforcement ($T{=}0.5$) a parent is
non-blocking once at least half of its sub-criteria pass. We define the
\emph{DAG gap} as the reduction in cumulative weighted Task Pass Rate
when DAG blocking is applied to the same outputs that were also
evaluated under full evaluation, and we measure this gap on every
(model, framework) configuration in
Table~\ref{tab:dag_gap}.

\begin{table*}[t]
  \centering
  \small
  \caption{DAG-gap analysis of cumulative weighted Task Pass Rate after Round~3, comparing strict ($T{=}1.0$, every parent failure blocks all children) and soft-satisfaction ($T{=}0.5$, a parent is non-blocking once at least 50\% of its sub-criteria pass) variants of the DAG-aware protocol against full evaluation. \textbf{Full}: weighted Task Pass Rate when every task is evaluated regardless of dependencies. \textbf{Gap\textsubscript{T}}: reduction (in pp) introduced by DAG blocking at threshold $T$. Mean gaps shown in the bottom row.}
  \label{tab:dag_gap}
  \setlength{\tabcolsep}{6pt}
  \begin{tabular}{ll | r | rr}
  \toprule
  \textbf{Framework} & \textbf{Model} & \textbf{Full (\%)} & \textbf{Gap\textsubscript{$T{=}1.0$}} & \textbf{Gap\textsubscript{$T{=}0.5$}} \\
  \midrule
  \multirow{6}{*}{Claude Code} & GPT-5.4 & 88.6 & 4.0 & 0.5 \\
   & Claude-4.6-Sonnet & 89.4 & 3.4 & 0.0 \\
   & GLM-5 & 77.9 & 3.8 & 1.2 \\
   & Qwen3.5-Plus & 70.8 & 8.4 & 0.7 \\
   & MiniMax-M2.7 & 77.2 & 9.9 & 0.8 \\
   & Seed-2.0-Pro & 53.4 & 9.3 & 1.6 \\
  \midrule
  \multirow{7}{*}{OpenHands} & GPT-5.4 & 90.1 & 3.6 & 0.0 \\
   & Gemini-3.1-Pro & 80.7 & 4.6 & 0.1 \\
   & GLM-5 & 76.0 & 7.6 & 0.8 \\
   & Kimi-K2.6 & 86.8 & 6.3 & 0.3 \\
   & Qwen3.5-Plus & 73.6 & 5.0 & 0.8 \\
   & MiniMax-M2.7 & 71.4 & 4.7 & 0.8 \\
   & Seed-2.0-Pro & 54.6 & 6.9 & 0.9 \\
  \midrule
  \multicolumn{2}{l|}{\textbf{Mean}} & --- & \textbf{6.0} & \textbf{0.6} \\
  \bottomrule
  \end{tabular}
\end{table*}

The strict variant produces a substantial gap (mean 6.0\,pp; 3.4--9.9\,pp
per row), since a parent that fails on a single sub-criterion is forced
to block its entire subtree even when the failure is unrelated to the
child's data contract. The soft-satisfaction variant we adopt
($T{=}0.5$) shrinks the mean gap to 0.6\,pp (worst case 1.6\,pp),
indicating that dependency enforcement under our protocol introduces
negligible distortion to the final scores while still preserving the
root-cause focus of the protocol. The tenfold reduction supports our
choice of $T{=}0.5$ as the default soft-satisfaction threshold.

\paragraph{DAG vs.\ Flat protocol.}
Table~\ref{tab:rq3_dag_vs_flat} compares the DAG-based evaluation
protocol against a Flat baseline that evaluates every criterion
independently. DAG numbers are recomputed under our adopted
$T{=}0.5$ soft-satisfaction protocol; Flat numbers and the evaluation
token totals are reported as in the previous version of this work
(re-running the Flat protocol was not necessary, because its score is
upper-bounded by the Full column of
Table~\ref{tab:dag_gap} and the difference relative to DAG at $T{=}0.5$
is small).

\begin{table*}[t]
  \centering
  \small
  \caption{Comparison of DAG-based vs.\ Flat evaluation protocols under the Claude Code framework (3 rounds). \textbf{Proj.}: Project Completion Rate; \textbf{Task\textsubscript{w}}/\textbf{Crit.\textsubscript{w}}: cumulative weighted pass rates after Round~3; \textbf{Eval Tokens}: total evaluation tokens consumed (input + output, in millions). DAG numbers are computed under the DAG-aware protocol with $T{=}0.5$ soft-satisfaction; Flat numbers are reported as in the previous version of this work, where every criterion is evaluated independently regardless of dependency status. DAG reduces evaluation token consumption by 23--27\% while achieving higher fix rates due to focused root-cause feedback.}
  \label{tab:rq3_dag_vs_flat}
  \setlength{\tabcolsep}{4pt}
  \begin{tabular}{ll|ccc|cc|cc|r}
  \toprule
  \multirow{2}{*}{\textbf{Protocol}} & \multirow{2}{*}{\textbf{Model}} & \multicolumn{3}{c|}{\textbf{Pass Rates (\%)}} & \multicolumn{2}{c|}{\textbf{Fix Rate R2 (\%)}} & \multicolumn{2}{c|}{\textbf{Fix Rate R3 (\%)}} & \textbf{Eval Tokens} \\
  & & Proj. & Task\textsubscript{w} & Crit.\textsubscript{w} & Task & Crit. & Task & Crit. & (M) \\
  \midrule
  \multirow{4}{*}{DAG} & GLM-5 & 30 & 76.6 & 82.9 & 61.8 & 66.6 & 44.7 & 48.7 & 17.9 \\
   & Seed-2.0-Pro & 8 & 51.8 & 60.9 & 35.7 & 38.7 & 35.4 & 42.9 & 21.9 \\
   & MiniMax-M2.7 & 22 & 76.4 & 86.3 & 57.7 & 68.0 & 35.4 & 47.8 & 21.2 \\
   & Qwen3.5-Plus & 24 & 70.1 & 80.2 & 44.4 & 51.8 & 31.4 & 38.2 & 18.4 \\
  \midrule
  \multirow{4}{*}{Flat} & GLM-5 & 36 & 83.6 & 89.0 & 45.9 & 49.0 & 38.7 & 46.6 & 23.4 \\
   & Seed-2.0-Pro & 16 & 69.6 & 79.4 & 30.9 & 35.0 & 28.1 & 39.9 & 26.9 \\
   & MiniMax-M2.7 & 26 & 74.0 & 84.9 & 39.9 & 43.3 & 27.9 & 40.4 & 25.7 \\
   & Qwen3.5-Plus & 18 & 77.3 & 86.6 & 40.8 & 45.3 & 26.9 & 34.0 & 25.0 \\
  \bottomrule
  \end{tabular}
\end{table*}

DAG yields slightly lower cumulative pass rates than Flat (e.g., GLM-5:
76.6\% vs.\ 83.6\% weighted Task Pass Rate), because tasks whose
prerequisites are unmet are not credited even when their own code
happens to be correct---an intentional consequence of dependency
enforcement, not an evaluation bias. In return, DAG is consistently
more efficient: it cuts evaluation token consumption by 23--26\%
(GLM-5: 17.9M vs.\ 23.4M tokens; Qwen3.5-Plus: 18.4M vs.\ 25.0M) and
yields uniformly higher fix rates because feedback is concentrated on
root-cause failures (e.g., GLM-5 R$_2$ task fix rate: 61.8\% vs.\
45.9\%; MiniMax R$_2$: 57.7\% vs.\ 39.9\%; Seed-2.0-Pro R$_2$: 35.7\%
vs.\ 30.9\%). Across all four models and both rounds the DAG protocol
posts a higher fix rate than Flat, supporting DAG as the preferred
protocol: \emph{more focused, slightly stricter, and substantially
cheaper}.

\section{Framework Comparison}
\label{sec:appendix_framework}
Across the five LLMs evaluated under both frameworks (GPT-5.4, GLM-5, Qwen3.5-Plus, MiniMax-M2.7, Seed-2.0-Pro), OpenHands posts a higher Project Completion Rate on every backbone (5-model mean 29.2\% vs.\ 25.6\%, $+3.6$\,pp), yet the cumulative weighted Task Pass Rate is essentially tied (72.5\% vs.\ 72.6\%) and Claude Code even leads on MiniMax-M2.7 (76.4\% vs.\ 70.7\%). The divergence indicates that Claude Code's failures are predominantly ``near-miss'' projects missing only one or two tasks out of fifteen-plus, which the strict all-pass PCR threshold amplifies into binary failures. Stronger first-round code on the top model contributes additionally: on GPT-5.4, OpenHands reaches 56.6\% Round-1 weighted Task Pass Rate versus 49.7\% on Claude Code, providing a higher baseline for subsequent refinement.

Despite this PCR gap, Claude Code is roughly $1.8\times$ faster end-to-end (64.5 vs.\ 116.3 minutes per project across three rounds) and uses tokens more efficiently for models that exercise plan mode (GPT-5.4 input tokens: 5.5M vs.\ 11.0M). The frameworks also fail in distinct ways: Claude Code mostly through near-miss tasks and infrastructure timeouts, OpenHands more often through process-management deadlocks where the agent fails to terminate its dev server or inadvertently kills the orchestrator via broad \texttt{pkill} commands.

\section{Implementation Details}
\label{sec:appendix_impl}
This appendix consolidates the operational details of our experiments, complementing the high-level setup in~\autoref{subsection:settings}.

\paragraph{Model--Framework Coverage.}
Three (model, framework) cells are skipped in Table~\ref{tab:main_results}:
\begin{itemize}[leftmargin=12pt,itemsep=1pt,topsep=2pt]
    \item \textbf{Gemini-3.1-Pro on Claude Code.} Gemini's native tool-call protocol is not supported by our in-house Claude Code sandbox; running it without protocol shimming produces malformed trajectories (cf.\ \url{https://github.com/coffeegrind123/gemini-for-claude-code}).
    \item \textbf{Kimi-K2.6 on Claude Code.} Kimi's reasoning protocol is not supported by our in-house Claude Code sandbox, leading to truncated rollouts in pilot.
    \item \textbf{Claude-4.6-Sonnet on OpenHands.} OpenHands' explicit-tool-call loop issues markedly more LLM calls per task than Claude Code; completing the full $50\,\text{tasks} \times 3\,\text{rounds}$ matrix at Sonnet's per-token rate exceeded our evaluation API budget, so we report Sonnet on Claude Code only.
\end{itemize}

\paragraph{Evaluation Hyperparameters.}
Unless otherwise specified, all experiments use:
\begin{itemize}[leftmargin=12pt,itemsep=1pt,topsep=2pt]
    \item \textbf{Generation:} \texttt{max\_steps}=3 (Code Agent rounds); temperature follows each model's default for tool-using agents.
    \item \textbf{Evaluation:} \texttt{max\_retry}=3 per dependency layer, browser concurrency $=3$, headless Chrome with vision-enabled UI Agent (max 100 navigation steps per criterion, 3600\,s wall-clock timeout). On the DAG, a parent task is marked as \emph{satisfied for downstream propagation} once at least 50\% of its sub-criteria pass (\emph{soft-satisfaction threshold} $T{=}0.5$); a parent's failure on a peripheral sub-criterion (e.g., a corner-case validation rule unrelated to the child task's data contract) should not block evaluation of otherwise-independent descendants. Empirically, $T{=}0.5$ reduces the mean DAG-blocking gap from 6.0 pp to 0.6 pp relative to strict ($T{=}1.0$) propagation while preserving the root-cause focus of the protocol.
    \item \textbf{LLM roles:} GPT-5.4 serves as the backbone for both the UI Agent (criterion judging) and the User LLM (feedback synthesis), held fixed across all Code Agent configurations to isolate Code Agent capability.
\end{itemize}

\paragraph{Reproducibility.}
Each task is run once per (model, framework, round) cell to keep the total budget tractable, and we treat the 50 projects in each cell as i.i.d.\ samples. Within a project, the per-project metric is a $\{0,1\}$ indicator (PCR) or a weighted pass ratio in $[0,1]$ (weighted Task / Criteria Pass Rate, weighted by difficulty $w_i$); the 50 per-project values are averaged into the cell mean $\bar{x}$, and the 95\%~CI is the symmetric CLT normal-approximation interval $\bar{x} \pm 1.96 \cdot s/\sqrt{50}$ with $s$ the unbiased sample standard deviation, computed via \texttt{scipy.stats.norm.interval}; subscripts in Table~\ref{tab:main_results} report the half-width (e.g., $90.1_{\pm 4.7}$ denotes $[85.4,94.8]$), and we use CIs to support qualitative comparisons (overlapping vs.\ non-overlapping) rather than as exact probability statements since the normal approximation can degrade for PCR cells near the 0--1 boundary.

\section{Per-Task-Type Performance Breakdown}
\label{sec:appendix_bytype}

To understand where the residual failures concentrate, we decompose
the cumulative weighted Task Pass Rate after Round~3 by task type.
Recall from Section~3.2 that every task is annotated as one of three
types: \tPres{} verifies that a required UI element exists,
\tFunc{} validates whether an interactive feature behaves correctly,
and \tRob{} checks edge-case and invalid-input handling.

\begin{table*}[t]
  \centering
  \caption{Per-task-type breakdown of cumulative weighted Task Pass Rate after three rounds, under the DAG-aware evaluation protocol ($T{=}0.5$). Tasks are partitioned into Presence (element existence), Functionality (interactive behaviour), and Robustness (edge-case / invalid input handling). Per-project averages with within-project difficulty weighting. Highest values per framework block in \textbf{bold}.}
  \label{tab:bytype_results}
  \setlength{\tabcolsep}{6pt}
  \begin{tabular}{ll ccc}
  \toprule
  \textbf{Framework} & \textbf{Model} & \textbf{Pres.\ (\%)} & \textbf{Func.\ (\%)} & \textbf{Rob.\ (\%)} \\
  \midrule
  \multirow{7}{*}{OpenHands} & GPT-5.4 & \textbf{98.0} & \textbf{90.1} & \textbf{86.0} \\
   & Kimi-K2.6 & 90.1 & 87.3 & 81.7 \\
   & Gemini-3.1-Pro & 95.8 & 81.3 & 71.5 \\
   & GLM-5 & 88.6 & 76.5 & 67.1 \\
   & Qwen3.5-Plus & 86.9 & 73.3 & 67.5 \\
   & MiniMax-M2.7 & 85.7 & 71.1 & 64.5 \\
   & Seed-2.0-Pro & 68.1 & 55.0 & 42.2 \\
  \midrule
  \multirow{6}{*}{Claude Code} & Claude-4.6-Sonnet & 93.4 & \textbf{89.5} & \textbf{87.6} \\
   & GPT-5.4 & \textbf{95.9} & 88.3 & 84.7 \\
   & GLM-5 & 88.0 & 78.1 & 70.6 \\
   & MiniMax-M2.7 & 92.9 & 77.8 & 66.2 \\
   & Qwen3.5-Plus & 85.4 & 71.3 & 62.7 \\
   & Seed-2.0-Pro & 71.0 & 52.8 & 45.9 \\
  \bottomrule
  \end{tabular}
\end{table*}

The ordering Presence~$>$~Functionality~$>$~Robustness holds for every
configuration in Table~\ref{tab:bytype_results}: rendering the right
elements is by far the easiest sub-problem (68.1--98.0\%), getting
their interactive behaviour correct is harder (52.8--90.1\%), and
handling edge cases is the hardest (42.2--87.6\%). The inter-model
spread also widens monotonically, from 30\,pp on Presence to 45\,pp on
Robustness, which means that Robustness is the dimension on which
stronger and weaker code agents separate most cleanly. Claude-4.6-Sonnet
on Claude Code is the only configuration that nearly closes the
Functionality--Robustness gap (89.5\% vs.\ 87.6\%), suggesting that
robustness on edge cases is a property of model training rather than
of agent scaffolding.

\section{Per-Round Fix Rate}
\label{sec:appendix_fix_rate}

Section~5 (RQ2) argues that initial generation and feedback-driven
repair are largely decoupled capabilities, and uses
Figure~\ref{fig:rq2_bar} to make this argument visually. For a
quantitative companion, Table~\ref{tab:fix_rate} reports the per-round
fix rate, defined as the fraction of previously-failed items that
become passing after the next round of refinement. Concretely, at the
task level
\[
  \text{FixRate}^{\text{task}}_s
  \;=\;
  \frac{|F_{s-1} \cap P_s|}{|F_{s-1}|},
\]
where $F_{s-1}$ is the set of tasks that appeared in the
\texttt{fail[]} list of round $s{-}1$ and $P_s$ is the set of tasks
that appeared in the \texttt{pass[]} list of round $s$. The criteria
variant counts at the sub-criterion level: a previously-failed
sub-criterion is fixed if its parent task fully passes in round $s$,
or if its sub-criterion text is judged \textit{pass} in round $s$.
Numerator and denominator are aggregated across all 50 projects and
reported as a single global ratio per round, matching the original
fix-rate convention from earlier work and ensuring that each task or
sub-criterion is counted at most once per round.

\begin{table*}[t]
  \centering
  \caption{Per-round fix rates. \textbf{Task fix rate at $R_s$} = $|F_{s-1} \cap P_s| / |F_{s-1}|$, where $F_{s-1}$ is the set of tasks that were in the \texttt{fail[]} list of round $s{-}1$ and $P_s$ is the set of tasks that passed in round~$s$. \textbf{Criteria fix rate} is analogous at the sub-criterion level: a previously-failed sub-criterion is counted as fixed if its parent task is fully passed in round~$s$ or if its sub-criterion text is judged \textit{pass} in round~$s$. Counts are aggregated across all 50 projects as a single global ratio. \textbf{The denominator differs across rows} (each row's prior-failure set depends on its own first-round performance), so column-wise ranking should be interpreted with care.}
  \label{tab:fix_rate}
  \small
  \setlength{\tabcolsep}{6pt}
  \begin{tabular}{ll cc cc}
  \toprule
  \multirow{2}{*}{\textbf{Framework}} & \multirow{2}{*}{\textbf{Model}} & \multicolumn{2}{c}{\textbf{Task Fix Rate (\%)}} & \multicolumn{2}{c}{\textbf{Criteria Fix Rate (\%)}} \\
  \cmidrule(lr){3-4} \cmidrule(lr){5-6}
  & & R$_2$ & R$_3$ & R$_2$ & R$_3$ \\
  \midrule
  \multirow{7}{*}{OpenHands} & GPT-5.4 & 62.6 & 52.6 & 71.2 & 63.7 \\
   & Kimi-K2.6 & 66.0 & 58.1 & 73.7 & 69.8 \\
   & Gemini-3.1-Pro & 61.2 & 43.6 & 70.1 & 53.9 \\
   & GLM-5 & 56.3 & 45.0 & 60.1 & 55.9 \\
   & Qwen3.5-Plus & 45.8 & 39.4 & 50.1 & 47.8 \\
   & MiniMax-M2.7 & 48.4 & 40.9 & 57.8 & 51.0 \\
   & Seed-2.0-Pro & 39.0 & 26.7 & 47.0 & 31.6 \\
  \midrule
  \multirow{6}{*}{Claude Code} & Claude-4.6-Sonnet & 67.9 & 52.2 & 72.7 & 63.0 \\
   & GPT-5.4 & 68.7 & 53.3 & 78.3 & 58.6 \\
   & GLM-5 & 61.8 & 44.7 & 66.6 & 48.7 \\
   & MiniMax-M2.7 & 57.7 & 35.4 & 68.0 & 47.8 \\
   & Qwen3.5-Plus & 44.4 & 31.4 & 51.8 & 38.2 \\
   & Seed-2.0-Pro & 35.7 & 35.4 & 38.7 & 42.9 \\
  \bottomrule
  \end{tabular}
\end{table*}

A subtlety worth flagging is that the denominator $|F_{s-1}|$ differs
across rows: a model that already converted most tasks in Round~1 has
a smaller residual failure set to repair from, which can lower its
later-round fix rates even when its absolute performance is high.
Cross-row comparisons should therefore focus on the trajectory within
a row (e.g., R$_2$ versus R$_3$ for the same model) rather than on
the magnitude of any single cell.

Two trends are nonetheless robust. First, every model's R$_2$ task
fix rate exceeds its R$_3$, consistent with the diminishing-returns
shape of the cumulative pass-rate curves in
Figure~\ref{fig:rq2_bar}. Second, models with lower first-round Task
Pass Rates do not necessarily have lower fix rates: Kimi-K2.6 on
OpenHands posts an R$_2$ task fix rate of 66.0\% from a 44.3\%
first-round baseline, exceeding several models with stronger first
rounds, which is the quantitative analogue of the ``decoupled''
observation in RQ2.

\section{Human Annotation Protocol}
\label{sec:appendix_annotation}

\begin{figure*}[htbp]
    \centering
    \includegraphics[width=0.95\linewidth]{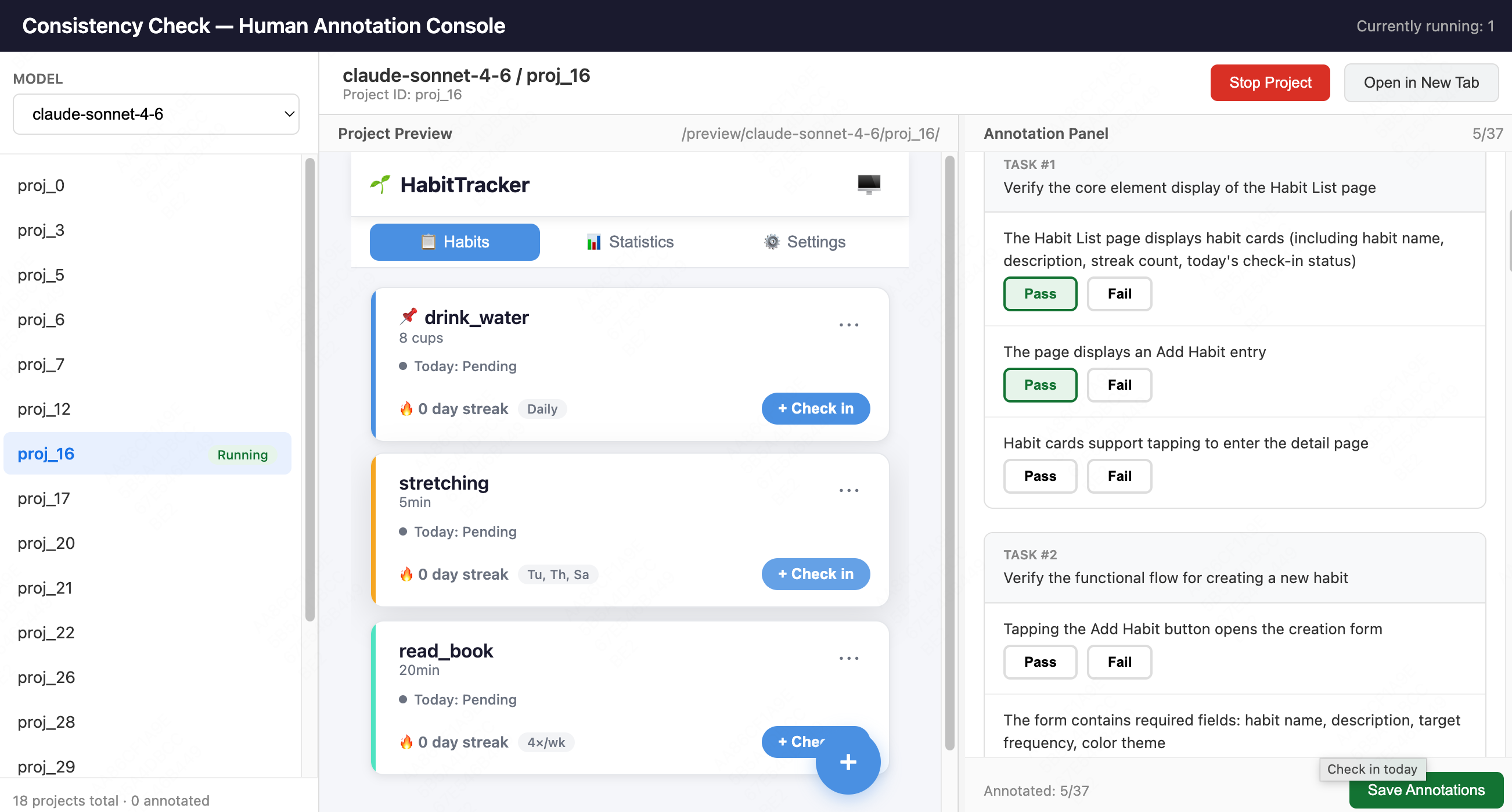}
    \caption{The web-based annotation console used by human annotators.
    The deployed project under evaluation is rendered live in the centre
    pane; the right pane lists the per-criterion checklist grouped by
    task; annotators interact with the live application and judge each
    criterion as Pass or Fail before saving.}
    \label{fig:annotation_ui}
\end{figure*}

This appendix records the rubric and conventions that human annotators
followed to produce the gold labels used in RQ3, together with the
annotation console (Figure~\ref{fig:annotation_ui}) used to apply
them. Annotation is \emph{single-annotated} (one independent annotator
per criterion) and applied only to the \emph{final-round} (Round~3)
deployed application; intermediate rounds are not annotated. Each
criterion is judged independently and the resulting JSON file follows
the exact same schema as our automated evaluator's
\texttt{eval\_step3\_results.json}, which makes the agreement
comparison schema-aligned by construction.

\begin{figure*}[htbp]
\begin{tcolorbox}[
    colback=gray!4,
    colframe=gray!55!black,
    boxrule=0.4pt,
    arc=2pt,
    left=8pt, right=8pt, top=6pt, bottom=6pt,
    fontupper=\small,
    title={\normalfont\sffamily\bfseries Annotation Rubric},
    coltitle=white,
    colbacktitle=gray!55!black,
]
\textbf{Pass criterion.}
A criterion is judged Pass if the behaviour it describes \emph{works
as expected} on the deployed application: the relevant UI element is
present and interactive, the interaction produces the expected
effect, and any associated data is correctly saved, updated, or
displayed.

\medskip
\textbf{Fail criterion.}
A criterion is judged Fail if any of the following hold:
\begin{itemize}[leftmargin=14pt,itemsep=1pt,topsep=2pt]
    \item \textbf{Missing functionality}: the UI element or feature
    described by the criterion is absent.
    \item \textbf{Incorrect behaviour}: the feature exists but the
    interaction outcome does not match the criterion's description.
    \item \textbf{Data error}: data is not correctly saved, updated,
    or displayed in the way the criterion specifies.
\end{itemize}

\medskip
\textbf{Conventions.}
Annotators were instructed to follow four conventions to keep
judgements aligned across the cohort:
\begin{enumerate}[leftmargin=14pt,itemsep=1pt,topsep=2pt]
    \item \textbf{Per-criterion independence.} Each criterion is
    judged on its own merits. A failing peer criterion within the
    same task does not propagate: if the current criterion's
    description is satisfied, it is marked Pass.
    \item \textbf{Faithful to the description.} Annotators judge only
    against the literal text of the criterion; additional acceptance
    standards beyond what the criterion specifies must not be
    introduced.
    \item \textbf{Style differences are not failures.} Cosmetic
    differences (colour, typography, spacing, exact wording of static
    labels) do not count against a criterion as long as the described
    functionality is correct.
    \item \textbf{Completeness.} Every criterion in the project must
    receive a Pass or Fail label before the annotation file is saved;
    skipped criteria are not allowed.
\end{enumerate}
\end{tcolorbox}
\caption{Rule for human annotators.}
\end{figure*}

\section{Prompt used to synthesize the underspecified user query from a Clarified PRD}
\label{sec:synthesize}
See~\autoref{fig:prompt_ambiguous_query}.
\begin{figure*}[htbp]
\begin{tcolorbox}[
    colback=gray!4,
    colframe=gray!55!black,
    boxrule=0.4pt,
    arc=2pt,
    left=6pt, right=6pt, top=4pt, bottom=4pt,
    fontupper=\small,
    title={\normalfont\sffamily\bfseries Prompt for Generating the Underspecified User Query from a Clarified PRD},
    coltitle=white,
    colbacktitle=gray!55!black,
]
\textbf{\# Task}\\[2pt]
You are a senior product manager responsible for generating an ambiguous user query for a Web frontend page based on its detailed Product Requirements Document (PRD). The ambiguous query will be used to instruct a Code Agent to develop the corresponding Web frontend project.

\begin{itemize}[leftmargin=12pt,itemsep=1pt,topsep=2pt]
    \item The ambiguous query should cover all functional requirements in the PRD, ensuring every functional point is reflected in the query.
    \item The query should be deliberately vague, avoiding overly specific technical terms or implementation details, so as to test the Code Agent's understanding and reasoning ability.
    \item The query should be written in natural language, with clear and easily understandable phrasing.
    \item The query must be \textbf{concise}: only state what the page needs to do, without elaborating on detailed requirements or implementation approaches. \textbf{Use as few words as possible to convey as much information as possible.}
    \item The query must not contain any information unrelated to the PRD; every description must be directly tied to the PRD.
    \item The query should avoid absolute wording such as ``must'' or ``necessarily'', in order to preserve ambiguity and flexibility.
    \item The query should be user-experience oriented, reflecting the needs and expectations of the end user when interacting with the page.
    \item The query should include necessary domain information such as label names or rule names, but avoid overly specific implementation details.
    \item Excluding the additional information (label names, rule names, etc.), the body of the ambiguous query must be kept within \textbf{200 words}.
\end{itemize}

\textbf{\# Example underspecified Query}\\[2pt]
Please build a stock analysis and customized report generation website for individual investors. The site should quantitatively score and comprehensively evaluate stocks based on market-data and fundamental indicators, generate a ranked leaderboard from high to low, and use a radar chart to present each stock's score structure across sub-indicator dimensions, so that strengths and weaknesses can be quickly identified.

\smallskip
\textit{Requirements:} indicators and scoring rules should be maintainable; after selecting a trading day or time window, market data is loaded and annotated with source and update time; market / fundamental / composite scores are automatically computed and a stable ranking is produced; sorting, searching, and filtering by industry / market are supported; clicking on a stock shows its radar chart and sub-indicator details, and allows one-click generation of a customized stock report; multi-stock comparison is supported; for missing or anomalous data, identification, reasons, and placeholder prompts should be shown.

\smallskip
\textbf{\# PRD content}\\[2pt]
\texttt{\{prd\}}
\end{tcolorbox}
\caption{Prompt used to synthesize the underspecified user query from a Clarified PRD.}
\label{fig:prompt_ambiguous_query}
\end{figure*}

\section{A Fully Compliant Implementation}
\label{sec:fulldata}

To calibrate what a successful Asuka-Bench task looks like
end-to-end, we walk through one project drawn from the dataset,
\texttt{proj\_34} -- \emph{Supplier Resource Evaluation Visualization
Ranking Page}.  This project is representative of the data-visualization
slice of Asuka-Bench: it requires a maintainable indicator system, a
weighted scoring engine, a sortable / filterable ranking list, a radar
chart with hover and multi-supplier comparison, and explicit fault
tolerance for missing or anomalous data.  Below we present, in order,
(i)~the underspecified user query the Code Agent receives; (ii)~the
Clarified PRD that defines ground truth; (iii)~the 16-task evaluation
rubric (Table~\ref{tab:case0_rubric}); (iv)~the task-dependency DAG
(Figure~\ref{fig:case0_dag}); and (v)~screenshots of the resulting page
(Figure~\ref{fig:case0_screens}).

\subsection{1.~Underspecified User Query}
See~\autoref{fig:Rawuserinput}.

\subsection{2.~Clarified PRD}
See~\autoref{fig:PRD}.

\subsection{3.~Evaluation Rubric}
See~\autoref{tab:case0_rubric}.

\begin{figure*}[htbp]
\begin{tcolorbox}[colback=gray!5,colframe=gray!50,boxrule=0.4pt,arc=2pt,
                 left=6pt,right=6pt,top=4pt,bottom=4pt,
                 title=\textbf{Raw user input (vague)}]
\small
\textbf{Supplier Resource Evaluation Visualization Ranking Page.}
This project builds a visualization ranking page for supplier
management and procurement evaluation scenarios.  It quantitatively
scores and comprehensively ranks suppliers based on resource-type and
quality-type indicators, displays the ranking from high to low, and
presents each supplier's score structure across sub-indicator
dimensions via a radar chart so that strengths and weaknesses can be
identified quickly.

\smallskip
\textbf{Loose requirements.} Maintainable indicators and scoring rules;
load data after selecting a period / scope and annotate source and
update time; automatically compute resource / quality / composite
scores and emit a stable ranking; support sort, search, and category /
region filtering; click a supplier to see its radar chart and
sub-indicator detail; support multi-supplier comparison; for missing
or anomalous data, show identification, reasons, and placeholder
prompts.
\begin{itemize}\setlength\itemsep{1pt}
\item \textbf{Indicator schema}: name $|$ type (resource / quality) $|$ scoring direction (positive / negative) $|$ weight $|$ belonging dimension.
\item \textbf{Rule schema}: normalization range $|$ threshold interval $|$ reverse-scoring strategy $|$ missing-data strategy.
\item \textbf{Ranking columns}: rank $|$ supplier name $|$ composite score $|$ resource score $|$ quality score.
\item \textbf{Radar tooltip}: dimension score $|$ sub-indicator list $|$ key raw values.
\item \textbf{Anomaly info}: anomalous field + supplier name $|$ entry point for non-evaluable reason.
\end{itemize}
\end{tcolorbox}
\caption{Vague user input of \texttt{proj\_34}}

\label{fig:Rawuserinput}
\end{figure*}

\begin{figure*}[htbp]
\begin{tcolorbox}[colback=blue!2,colframe=blue!40,boxrule=0.4pt,arc=2pt,
                 left=6pt,right=6pt,top=4pt,bottom=4pt,
                 title=\textbf{PRD --- Supplier Resource Evaluation Visualization Ranking Page}]
\small
\textbf{1.~Project Overview.}
A visualization ranking page for supplier management and procurement
evaluation.  Suppliers are scored on resource-type and quality-type
indicators and ranked by composite score; a radar chart shows each
supplier's per-dimension score structure for quick strength /
weakness diagnosis.

\smallskip
\textbf{2.~Core Functional Requirements.}

\textit{2.1~Indicator System \& Scoring Rules.}
Maintain an indicator list with the fields: \emph{name, type, scoring
direction, weight, belonging dimension}.  Resource-type and
quality-type indicators must be scored under their respective rule
configurations: \emph{normalization range, threshold intervals,
reverse-scoring strategy}.

\smallskip
\textit{2.2~Data Loading \& Scope Selection.}
A statistical-period selector and data-scope selector control loading;
switching the period synchronously refreshes the ranking and the
visualization.  Each indicator's raw data must be field-matched against
the indicator list, and data update time and source identifiers
(e.g.\ ERP, QMS) must be displayed for traceability.

\smallskip
\textit{2.3~Score Calculation \& Ranking Generation.}
For each supplier, compute per-indicator scores, the resource sub-score,
the quality sub-score, and a weighted composite score; positive- and
negative-direction indicators are scored independently and mapped onto
the same scale before aggregation.  The ranking is sorted by composite
score in descending order; ties are broken by a fixed, stable rule.

\smallskip
\textit{2.4~Composite Ranking Display.}
The ranking list contains, at minimum, columns for \emph{rank, supplier
name, composite score, resource score, quality score}; default sort is
by composite score, and column headers (resource, quality, any
sub-indicator) toggle alternative sort keys.  The list supports search
by supplier name and filters on supplier category and region.

\smallskip
\textit{2.5~Sub-Indicator Detail \& Radar Chart.}
Selecting a supplier populates that supplier's radar chart, whose axes
correspond to the indicator's belonging dimensions or a preset
dimension set.  Hover tooltips show \emph{dimension score, the
sub-indicator list under that dimension, and key raw values}.
A sub-indicator detail list (name, type, raw value, score, weight,
score contribution) accompanies the radar chart.
\textbf{Comparison mode allows simultaneous display of multiple
suppliers' dimension curves on the same radar chart.}

\smallskip
\textbf{3.~Fault Tolerance \& Exception Handling.}

\textit{3.1~Missing / Incomplete Data.}
When raw indicator data is missing, apply the configured missing
strategy (mark-missing or median-imputation) and clearly mark missing
items in the list and detail view.  \textbf{If missing data makes the
composite score uncomputable, mark the supplier as \emph{non-evaluable},
exclude it from the default ranking, and provide an entry point to
inspect the reason.}

\smallskip
\textit{3.2~Out-of-Range / Non-Numeric Input.}
When raw data exceeds the rule-allowed range or is non-numeric, block
scoring and prompt with the anomalous field and supplier name; the
supplier shows an error-state badge in the list.

\smallskip
\textit{3.3~Radar-Chart Robustness.}
On division-by-zero, infinity, or empty dimensions that would make the
radar chart unrenderable, fall back to a placeholder state and prompt
the user to adjust the data or rule configuration.

\smallskip
\textbf{4.~Appearance.}
Dashboard-style layout with a clear hierarchy and a distinct
ranking-area / detail-area separation.  Resource-type and quality-type
information must be visually distinguished consistently across the
whole page.
\end{tcolorbox}

\caption{Full PRD doc of \texttt{proj\_34}}

\label{fig:PRD}
\end{figure*}

\begin{table*}[ht]
\centering
\scriptsize
\setlength{\tabcolsep}{5pt}
\renewcommand{\arraystretch}{1.30}
\begin{tabular}{@{}
  >{\centering\arraybackslash}m{0.030\linewidth}
  >{\raggedright\arraybackslash}m{0.165\linewidth}
  >{\centering\arraybackslash}m{0.105\linewidth}
  >{\centering\arraybackslash}m{0.030\linewidth}
  >{\raggedright\arraybackslash}m{0.610\linewidth}
@{}}
\toprule
\# & Task summary & Type & Wt. & Criteria \\
\midrule
0 & Page-layout elements present & \tPres &2 &
\cn{1} Statistical-period selector at the top. \cn{2} Supplier ranking list container in the main area. \cn{3} Radar-chart container in the detail area. \cn{4} Supplier-name search box and category / region filter dropdowns. \\

1 & Period switching & \tFunc &3 &
\cn{1} Switching the period changes the ranking-list data. \cn{2} Radar chart and detail data update synchronously. \cn{3} An empty period shows an empty-state prompt. \\

2 & Ranking columns \& consistency & \tFunc &3 &
\cn{1} Columns include rank, supplier name, composite, resource, and quality scores. \cn{2} List data matches the selected period. \cn{3} All three scores are displayed numerically. \\

3 & Sortable list & \tFunc &4 &
\cn{1} Default sort is by composite score, descending. \cn{2} Clicking the resource header sorts by resource score. \cn{3} Clicking the quality header sorts by quality score. \cn{4} Clicking any sub-indicator header sorts by that sub-indicator. \\

4 & Search \& filter intersection & \tFunc &3 &
\cn{1} Name-keyword search filters the list. \cn{2} Category filter narrows to that category. \cn{3} Region filter narrows to that region. \cn{4} Combined search + filter returns the intersection. \\

5 & \textbf{Resource-type scoring} & \tFunc &5 &
\cn{1} Resource sub-indicator detail shows raw value and score. \cn{2} Resource sub-score equals the weighted aggregate of its sub-indicator scores. \cn{3} For negative indicators (e.g.\ Avg.\ Lead Time), smaller raw values yield higher scores. \\

6 & \textbf{Quality-type scoring} & \tFunc &5 &
\cn{1} Quality sub-indicator detail shows raw value and score. \cn{2} Quality sub-score equals the weighted aggregate of its sub-indicator scores. \cn{3} For negative indicators (e.g.\ Defect PPM), smaller raw values yield higher scores. \\

7 & Composite-score correctness & \tFunc &4 &
\cn{1} Composite equals the weighted combination of resource + quality scores. \cn{2} Numeric precision matches the display requirement. \cn{3} Composite shown in the list matches the value shown in the detail panel. \\

8 & Radar chart per supplier & \tFunc &4 &
\cn{1} Selecting a supplier populates that supplier's radar chart. \cn{2} Axes correspond to indicator-belonging dimensions (or the preset dimension set). \cn{3} Plotted points reflect each dimension's score. \\

9 & Radar hover tooltip & \tFunc &3 &
\cn{1} Hovering a dimension shows a tooltip. \cn{2} Tooltip displays the dimension score. \cn{3} Tooltip lists the sub-indicators under that dimension and their key raw values. \\

10 & Multi-supplier comparison & \tFunc &4 &
\cn{1} Selecting comparison mode plus multiple suppliers renders multiple curves. \cn{2} Curves are distinguished by color or legend. \cn{3} Hovering a dimension shows the comparison values across selected suppliers. \\

11 & Sub-indicator detail list & \tFunc &3 &
\cn{1} Detail area shows a sub-indicator list. \cn{2} Columns include name, type, raw value, score, weight, and score contribution. \cn{3} Data matches the selected supplier and period. \\

12 & Source \& update-time annotation & \tFunc &3 &
\cn{1} Page or detail area shows the data update time. \cn{2} Each indicator is annotated with its source (e.g.\ ERP, QMS). \cn{3} Source labels are consistent with the metadata configuration. \\

13 & \textbf{Missing-data handling} & \tRob &4 &
\cn{1} Missing raw values trigger the configured strategy (mark-missing / median-imputation). \cn{2} Missing items have a clear visual indicator in the list / detail. \cn{3} If composite is incomputable, the supplier is marked \emph{non-evaluable} and excluded from the default ranking. \\

14 & \textbf{Anomalous-input handling} & \tRob &4 &
\cn{1} Out-of-range raw values block scoring. \cn{2} Non-numeric inputs prompt with the anomalous field and supplier name. \cn{3} Anomalous suppliers carry an error-state badge in the list. \\

15 & \textbf{Radar-chart fallback} & \tRob &3 &
\cn{1} Infinite or uncomputable dimension scores must not crash the radar chart. \cn{2} If unrenderable, the area shows a placeholder state. \cn{3} The system prompts the user to check data / rule configuration. \\
\bottomrule
\end{tabular}
\caption{Evaluation rubric of \texttt{proj\_34}.
Tasks~\textbf{5} and~\textbf{6} (resource / quality scoring) carry the
heaviest weights; tasks~\textbf{13}, \textbf{14}, \textbf{15} are the
robustness frontier typically missed by single-pass generation.}
\label{tab:case0_rubric}
\end{table*}

\subsection{4.~Task-Dependency DAG}

\begin{figure*}[t]
\centering
\includegraphics[width=0.95\linewidth]{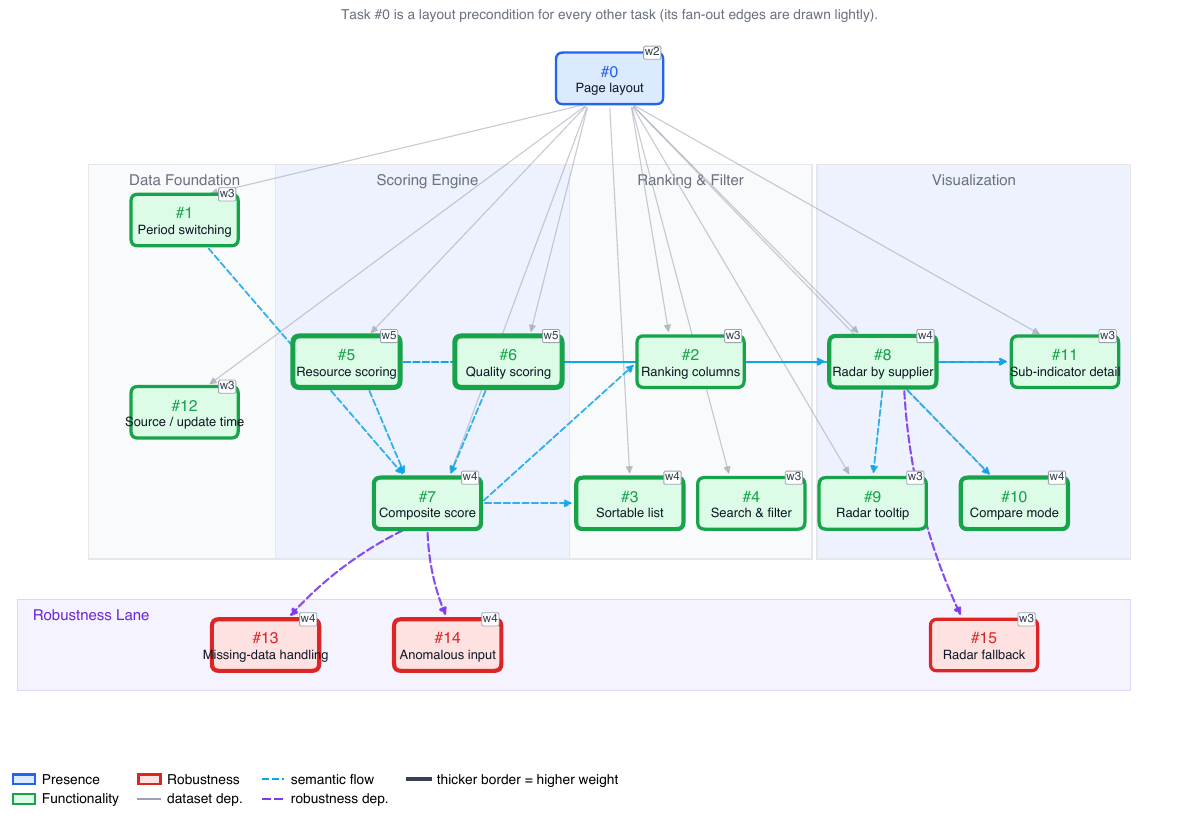}
\caption{Task-dependency DAG for the case study (\texttt{proj\_34},
16~tasks, total weight $=57$). Solid grey = dataset dependency
required by the DAG-aware scheduler; dashed sky-blue = semantic flow
that the implementation must honour even though the scheduler does
not enforce it; dashed purple = robustness gates entering the
bottom lane.  Border thickness encodes task weight (1\,$\to$\,5).}
\label{fig:case0_dag}
\end{figure*}

Figure~\ref{fig:case0_dag} renders the dependencies between the 16
tasks.  Filled colour encodes task type (\tPres / \tFunc / \tRob);
node-border thickness encodes weight; the four swim-lanes correspond
to the PRD's four functional areas (data foundation, scoring engine,
ranking \& filter, visualization), and the bottom lane gathers the
three robustness tasks.  The dataset only enforces ``everything depends
on the page-layout root'' plus three robustness gates
(\#13,~\#14~$\leftarrow$~\#7; \#15~$\leftarrow$~\#8); the dashed
sky-blue edges expose the \emph{semantic flow} a correct
implementation must additionally honour: scoring~$\to$~composite,
composite~$\to$~ranking, scored data~$\to$~radar, radar~$\to$~tooltip
and comparison.  These hidden edges are where most single-pass models
break.

\subsection{5.~Resulting Page (Successful Implementation)}

\begin{figure*}[t]
\centering
\includegraphics[width=\linewidth]{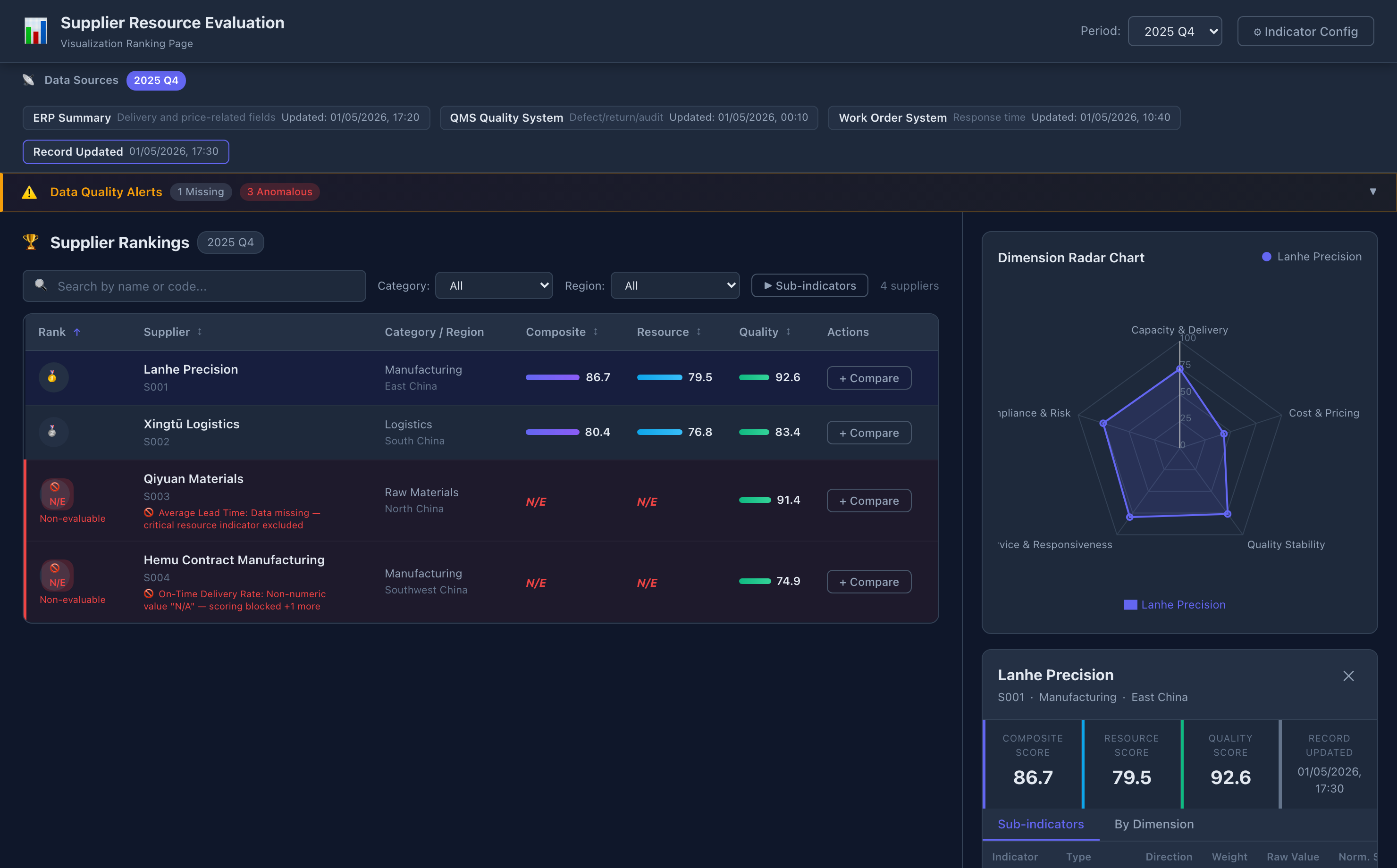}
\caption{Successful implementation of the case-study PRD by
Claude-4.6-Sonnet after three rounds of feedback.}
\label{fig:case0_screens}
\end{figure*}

Figure~\ref{fig:case0_screens} shows the final page produced by
Claude-4.6-Sonnet (Claude Code framework) after three rounds of
iterative repair (R1~6/16 $\to$~R2~11/16 $\to$~R3~14/16; the two
remaining failures are tasks~\#10 and~\#15).  The page implements every
PRD module: a header data-source bar with last-update timestamps; a
left ranking list with composite / resource / quality score progress
bars, sortable headers, search and category / region filters, and an
explicit \emph{non-evaluable} badge for suppliers excluded by the
missing-data rule; a right detail panel with the per-supplier radar
chart and the sub-indicator detail table.  Anomalous inputs surface
inline (e.g.\ ``On-Time Delivery Rate: Non-numeric value''), and the
data-quality alerts strip at the top exposes the system-level
robustness state.

\section{Case Study: Progressive Repair Through Iterative Feedback}
\label{sec:appendix_case_study}
Asuka-Bench's criterion-level scoring exposes failure modes that project-level pass/fail metrics hide. We illustrate this on two cases drawn from different UI families: a single-page analytics dashboard and a multi-page social app. Both show the same pattern under one round of low-verbosity feedback. Surface-rendering defects are fixed cleanly; deeper interaction-logic defects (sign-change reasoning in Case 1, file-reader wiring in Case 2) survive the round and would need richer feedback or another iteration to resolve. The split is only visible because every criterion is graded independently.

\begin{figure*}[htbp]
\centering
\includegraphics[width=\textwidth]{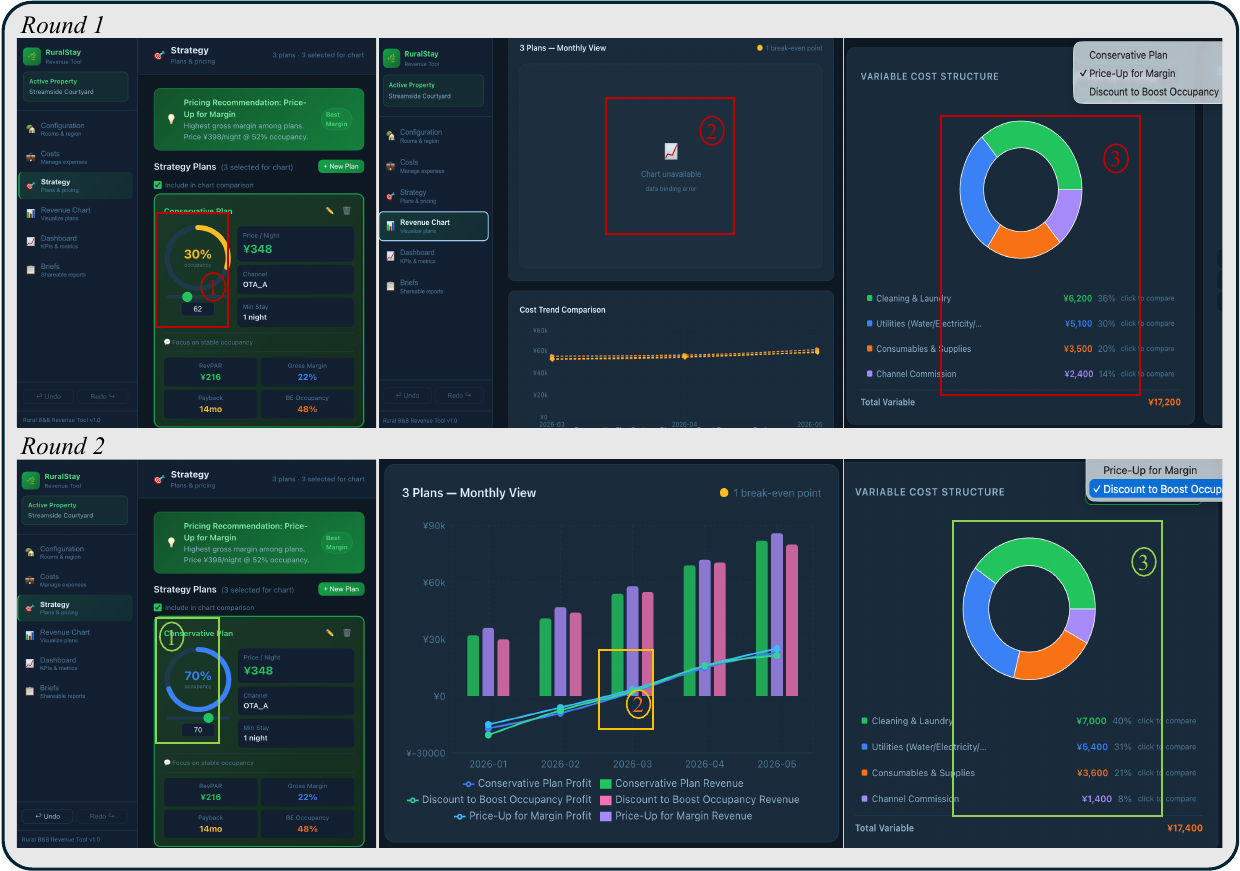}
\caption{Progressive repair on the Rural-Stay Revenue Strategy Tool (Claude-4.6-Sonnet). v1 (left) shows four highlighted failing regions; v2 (right) shows the post-Round-1 state with \textcolor{cFix}{\checkmark} fixed (green) and \textcolor{cFail}{\ding{55}} still failing (red). Pass count: v1 7/11 $\rightarrow$ v2 10/11 criteria.}
\label{fig:case_proj39}
\end{figure*}

\paragraph{Case 1: Rural-Stay Revenue Strategy Tool (Figure~\ref{fig:case_proj39}, Table~\ref{tab:case39_defects}).}
Claude-4.6-Sonnet generates a single-page revenue dashboard. v1 has four defects on the dial, the Revenue Chart, the break-even highlight, and the variable-cost pie. After Round-1, three (\textcircled{\scriptsize 1}, \textcircled{\scriptsize 2}, \textcircled{\scriptsize 4}) are cleanly fixed, while the break-even highlight (\textcircled{\scriptsize 3}), which requires sign-change reasoning on the profit curve, remains unfixed. Net: 7/11 $\rightarrow$ 10/11 criteria.

\begin{figure*}[htbp]
\centering
\includegraphics[width=\textwidth]{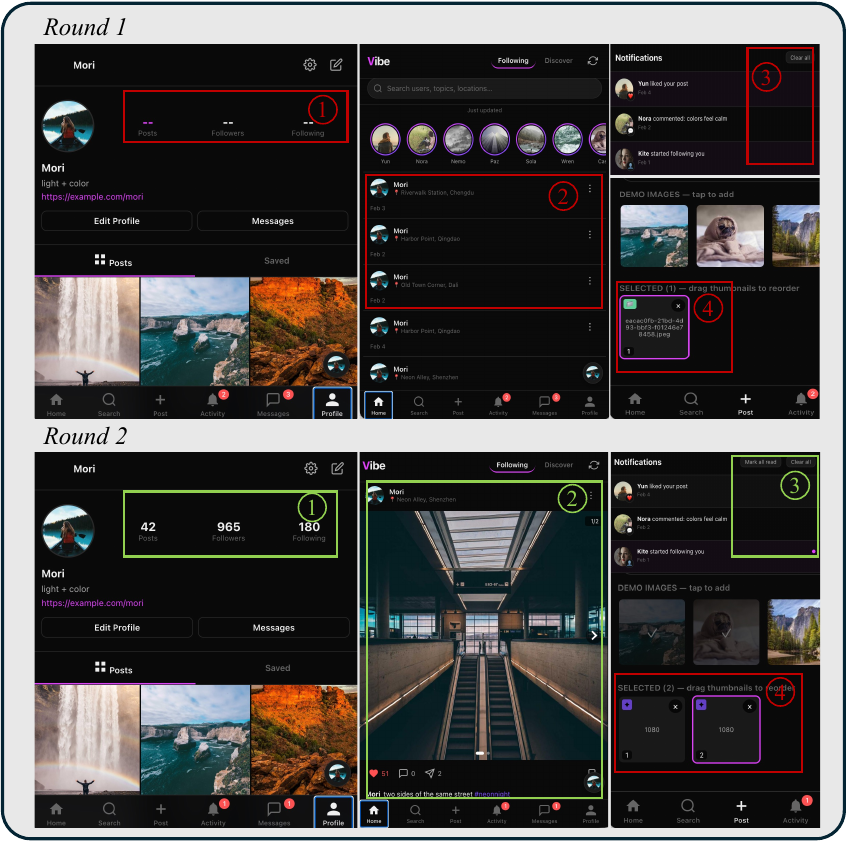}
\caption{Progressive repair on the Visual Social Media Platform (Claude-4.6-Sonnet). v1 (left) shows four cross-page failing regions; v2 (right) shows the post-Round-1 state with \textcolor{cFix}{\checkmark} fixed (green) and \textcolor{cFail}{\ding{55}} still failing (red). Pass count: v1 13/20 $\rightarrow$ v2 16/20 criteria.}
\label{fig:case_proj47}
\end{figure*}

\paragraph{Case 2: Visual Social Media Platform (Figure~\ref{fig:case_proj47}, Table~\ref{tab:case47_defects}).}
Claude-4.6-Sonnet generates a six-tab social platform. v1 has four cross-page rendering defects (\textcircled{\scriptsize 1}--\textcircled{\scriptsize 4}). After Round-1, three are fixed; only the Publish thumbnail (\textcircled{\scriptsize 3}) is misread as a label issue and left unfixed. Net: 13/20 $\rightarrow$ 16/20 criteria.

\begin{table*}[htpb]
\centering
\footnotesize
\setlength{\tabcolsep}{4pt}
\renewcommand{\arraystretch}{1.20}
\begin{tabular}{@{} c l >{\raggedright\arraybackslash}p{4.6cm} >{\raggedright\arraybackslash}p{4.6cm} @{}}
\toprule
\# & Region & R1 feedback (low-verbosity) & Repair verdict (v2) \\
\midrule
\textcircled{\scriptsize 1} & Dial & \emph{crit 3-0} — Dial drag does not update the numeric input. & \textcolor{cFix}{\checkmark\ fixed.} Drag$\to$input sync restored. \\
\textcircled{\scriptsize 2} & Revenue Chart & \emph{crit 4-0} — Chart panel shows a placeholder; curves not rendered. & \textcolor{cFix}{\checkmark\ fixed.} \texttt{ComposedChart} re-mounted on \texttt{chartData}. \\
\textcircled{\scriptsize 3} & Break-even mark & \emph{crit 5-0} — Zero-crossing of profit is not highlighted. & \textcolor{cFail}{\ding{55}\ unfixed.} Sign-change detection missing. \\
\textcircled{\scriptsize 4} & Cost pie & \emph{crit 6-1} — Pie does not respond to the plan dropdown. & \textcolor{cFix}{\checkmark\ fixed.} Filter key swapped to \texttt{selectedBreakdownPlan}. \\
\bottomrule
\end{tabular}
\caption{Per-defect summary for Case 1 (\emph{Rural-Stay Revenue Strategy Tool}).}
\label{tab:case39_defects}
\end{table*}

\begin{table*}[htbp]
\centering
\footnotesize
\setlength{\tabcolsep}{4pt}
\renewcommand{\arraystretch}{1.20}
\begin{tabular}{@{} c l >{\raggedright\arraybackslash}p{4.6cm} >{\raggedright\arraybackslash}p{4.6cm} @{}}
\toprule
\# & Page & R1 feedback (low-verbosity) & Repair verdict (v2) \\
\midrule
\textcircled{\scriptsize 1} & Profile & \emph{task 0 / crit 1} — Hero shows ``\texttt{--}'' for post/follower/following counts. & \textcolor{cFix}{\checkmark\ fixed.} Re-binds \texttt{user.stats.*} via \texttt{formatCount}. \\
\textcircled{\scriptsize 2} & Feed & \emph{task 2 / crit 2} — Feed card renders only the author header; media/caption/actions missing. & \textcolor{cFix}{\checkmark\ fixed.} Full PostCard body restored. \\
\textcircled{\scriptsize 3} & Publish & \emph{task 6 / crit 0} — After file selection, filename is shown instead of a thumbnail. & \textcolor{cFail}{\ding{55}\ unfixed.} \texttt{FileReader} wiring untouched. \\
\textcircled{\scriptsize 4} & Notifications & \emph{task 19 / crit 1} — Unread red-dot and header unread-count chip missing. & \textcolor{cFix}{\checkmark\ fixed.} Dot + chip re-added under \texttt{unread > 0}. \\
\bottomrule
\end{tabular}
\caption{Per-defect summary for Case 2 (\emph{Visual Social Media Platform}).}
\label{tab:case47_defects}
\end{table*}

\section{AI Usage Statement}
In the preparation of this manuscript, LLMs were used solely for grammatical refinement and phrasing adjustment of the text. All core intellectual contributions of this study---including the benchmark design (the underspecified-query / Clarified-PRD pairing, the criterion DAG, the soft-satisfaction protocol, and the closed-loop evaluation framework), the experimental setup (model and framework selection, metric definition, hyperparameter choices, and the human-agreement protocol), and all analysis and interpretation of results---were independently completed by the authors without any reliance on LLMs for idea generation, technical design, or result interpretation. All LLM-refined text has undergone manual review and revision by the authors to ensure alignment with the study's actual methods and results.

\end{document}